\documentclass{aa51}

\usepackage{graphicx}
\usepackage{amsmath,amsfonts,amssymb,amsxtra}
\usepackage{array}

\usepackage{natbib}
\bibpunct{(}{)}{;}{a}{}{,}

\newcommand{\rhor}{$\rho_\mathrm{R}$}
\newcommand{\rhow}{$\rho_\mathrm{W}$}
\newcommand{\rwd}{$R_\mathrm{WD}$}

\newcommand{\bdip}{$B^\mathrm{d}_\mathrm{pol}$} 
\newcommand{\bqua}{$B^\mathrm{q}_\mathrm{pol}$}
\newcommand{\thed}{$\Theta^\mathrm{d}$}
\newcommand{\phid}{$\Phi^\mathrm{d}$}
\newcommand{\theq}{$\Theta^\mathrm{q}$}
\newcommand{\phiq}{$\Phi^\mathrm{q}$}
\newcommand{\xoff}{${x'}_\mathrm{off}$}
\newcommand{\yoff}{${y'}_\mathrm{off}$}
\newcommand{\zoff}{${z'}_\mathrm{off}$}
\newcommand{\roff}{${\vec{r}'}_\mathrm{off}$}
\newcommand{\chisq}{$\chi^2$}
\newcommand{\chisqred}{$\chi^2_\mathrm{red}$}

\newcommand{\bpd}{\mbox{$B$--$\psi$ diagram}}

\begin{document}

\title{Zeeman tomography of magnetic white dwarfs}
\subtitle{I. Reconstruction of the field geometry from synthetic spectra}

   \author{F.~Euchner\inst{1} \and
        S.~Jordan\inst{1}\fnmsep\inst{2}\fnmsep\inst{3} \and
        K.~Beuermann\inst{1} \and
        B.T.~G\"ansicke\inst{1} \and
        F.V.~Hessman\inst{1}}

        \offprints{F.~Euchner, \email{feuchner@uni-sw.gwdg.de}}
   
        \institute{Universit\"ats-Sternwarte G\"ottingen, 
        Geismarlandstr.~11, D-37083 G\"ottingen, Germany
       	\and
        Institut f\"ur Astronomie und Astrophysik, 
        Universit\"at Kiel, D-24098 Kiel, Germany
	\and
        Institut f\"ur Astronomie und Astrophysik, 	
        Universit\"at T\"ubingen, Sand 1, D-72076 T\"ubingen, Germany}
             
\date{Received February 7, 2002 / Accepted May 15, 2002}

\abstract {We have computed optical Zeeman spectra of magnetic white
dwarfs for field strengths between 10 and 200\,MG and effective
temperatures between 8000 and 40\,000\,K. They form a database
containing 20\,628 sets of flux and circular polarization spectra.  A
least-squares optimization code based on an evolutionary strategy can
recover relatively complex magnetic field topologies from
phase-resolved synthetic Zeeman spectra of rotating magnetic white
dwarfs.  We consider dipole and quadrupole components which are
non-aligned and shifted off-centre. The model geometries include stars
with a single high-field spot and with two spots separated by
$\sim$90\degr.  The accuracy of the recovered field structure
increases with the signal-to-noise ratio of the input spectra and is
significantly improved if circular polarization spectra are included
in addition to flux spectra. We discuss the strategies proposed so far
to unravel the field geometries of magnetic white dwarfs.
 }
   
\titlerunning{Zeeman tomography of magnetic white dwarfs. I}
\authorrunning{F.~Euchner et al.}
\maketitle

\section{Introduction}

About 3\% of all white dwarfs have strong magnetic fields between
$10^6$ and $10^9$\,Gauss \citep{wickramasinghe+ferrario00-1,
jordan01-1}. In many of these magnetic white dwarfs (MWDs), the
surface field geometries deviate from simple centred dipoles. This
holds for isolated MWDs and for the MWDs in accreting close binaries
\citep{wickramasinghe+ferrario00-1,schwope95-1}.  While higher modes
are often thought to decay on timescales longer than the \mbox{$\tau
\ga 10^{10}$\,yr} of the fundamental mode, \citet{muslimovetal95-1}
showed that quadrupole or octupole components may survive via the Hall
effect if an internal toroidal magnetic field component is
present. Therefore, studies of the surface field structure provide
clues on the internal field configuration and its influence on the
evolution of MWDs.

The photospheric spectra of hydrogen-rich MWDs are characterized by
broad absorption structures formed by the overlap of numerous
components of the Balmer lines, shifted by hundreds or even thousands
of \AA\ from their zero-field positions by the linear and quadratic
Zeeman effects. These shifts completely obliterate the Doppler shifts
caused by rotation even in the most rapidly rotating MWDs.  As a
consequence, the Zeeman-Doppler imaging method devised for the
analysis of rapidly rotating magnetically active main sequence stars
\citep{semel89-1, donatietal89-1, brownetal91-1} is not applicable to
MWDs. The field geometry of MWDs can be derived, however, from the
analysis of the pure Zeeman splitting of the photospheric lines and
their circular polarization properties as a function of rotational
phase.  Because of the large Zeeman shifts, this approach must include
the whole optical range for \mbox{$B \ga 50$\,MG}. In the absence of
positional information from the Doppler effect, however, the inversion
of the flux and polarization spectra is an intricate task.
Trial-and-error fits of centred or shifted dipoles and quadrupoles
\citep{wickramasinghe+cropper88-1, putney+jordan95-1} are incapable of
exploring the full parameter space of possible solutions. We present a
new strategy using a pre-computed database of synthetic MWD spectra
and an automatic quality-of-fit optimization algorithm.

A first approach along these lines was presented by
\citet{donatietal94-1}, who used a maximum entropy algorithm (MEM) to
fit a matrix of areal filling factors for a grid of synthetic flux and
circular polarization spectra to simulated input data.  This way, they
obtained the `simplest' and, according to \emph{Occam's razor}, most
likely frequency distribution of transverse and longitudinal field
strengths over the visible stellar disc at each rotational phase, a
so-called ZEBRA plot, but this approach does not provide any
information about the spatial structure of the magnetic field.

In this paper, we investigate to what extent the underlying global
magnetic field distribution can be recovered directly from
least-squares fits to phase-resolved flux and polarization spectra of
a given signal-to-noise ratio. This approach uses the spatial
information provided by the magnetic fields seen at different
rotational phases and has the advantage that the uncertainties in the
parameters describing the global field structure can be directly
related to the noise in the spectra. Its disadvantage lies in the
necessary restriction to model fields which can be described by a
sufficiently small number of parameters.

We assume fields which can be represented by centred or offset dipole
and quadrupole components which need not be aligned with each
other. The specific geometries tested here include a star with a
single high-field spot and one with two spots separated by
$\sim$90\degr. Our computer code allows us to calculate areal filling
factor matrices analogous to ZEBRA plots, the resulting flux spectra,
and the wavelength-dependent circular polarization for a given
magnetic field model viewed at a number of rotational phases. We
compare the results with the reference input (which are simulated data
in this case) and determine the best-fit parameters using an
evolutionary optimization strategy.

The present paper is arranged as follows. In Sect.~\ref{sec:database}
we describe the database of flux and polarization spectra computed for
a wide range of field strengths, viewing angles, and effective
temperatures.  Section~\ref{sec:mfgeometry} describes the general
design of the magnetic field models and Sect.~\ref{sec:input_spectra}
the construction of the integrated spectra from the database for a
given model of the magnetic field. Section~\ref{sec:reconstruction}
explains the optimization code, describes the specific field models
subjected to the reconstruction tests, and investigates the ability of
the code to deduce the respective field parameters from the integrated
flux and polarization spectra. Finally, the power and also the
limitations of our approach are discussed in
Sect.~\ref{sec:discussion}.

In forthcoming papers, we will analyse phase-resolved spectral flux
and circular polarization data of MWDs obtained at the ESO VLT.

\section{The database}
\label{sec:database}

\subsection{Radiative transfer for Magnetic White Dwarfs}

Our synthetic Zeeman spectra and wavelength-dependent polarization
data are computed with the most recent version of the code developed
by S.~Jordan.  The polarization originates from the different
absorption coefficients $\kappa_\mathrm{l}$, $\kappa_\mathrm{r}$, and
$\kappa_\mathrm{p}$ for left- and right-handed circularly polarized
light, and linearly polarized light travelling perpendicularly to the
magnetic field, respectively, and is described by the four Stokes
parameters $I$, $Q$, $V$, and $U$.  The influence of the Faraday
rotation and the Voigt effect is accounted for by the magneto-optical
parameters \rhor\ and \rhow. The three radiative transport equations
of \citet{unno56-1} then expand into four equations
\citep{beckers69-1,hardorpetal76-1} which can be solved by one of
several different algorithms: (a) the method of
\citet{wickramasinghe+martin79-1} assumes that the source function is
linear in the optical depth and that between two successive depth
points the Stokes parameters can be described by exponential
functions; (b) direct Runge-Kutta integration; (c) accelerated
$\Lambda$ iterations \citep{takeda91-1}; (d) an approximation for
large Faraday rotation \citep{ramaty69-1}; or (e) matrix exponential
solutions \citep{dittmann95-1}.  Intensive tests performed by
H.~Schmidt and S.~Jordan in Kiel have demonstrated the numerical
equivalence of these methods with high accuracy. For the present
paper, we have calculated an extensive database of synthetic flux and
circular polarization spectra using Ramaty's approximation, which is
always justified in white dwarf atmospheres, and is rather efficient
with regard to CPU time.

The temperature and pressure structure of our atmospheres is taken
from zero field LTE models (\citealt{koesteretal79-1}). As a
consequence, the magnetic pressure and magnetic blanketing
(\citealp[][]{jordan92-1}) have been neglected. Convection is assumed
to be suppressed by the field \citep{jordan01-1}.  For the line
opacities, data from the T\"ubingen group
\citep{forsteretal84-1,roesneretal84-1,wunneretal85-1} were used.
Bound-free opacities were calculated using a modified approximation by
\citet{lamb+sutherland74-1} which leads to small errors only
\citep{jordan92-1, meranietal95-1, jordan+merani95-1} and saves an
enormous amount of computing time.  The approach described here was
developed in two diploma theses \citep{euchner98-1,rahn99-1} and was
also implemented by \citet{burleighetal99-1}.

\subsection{Database spectra}

\begin{figure*}[t]
\includegraphics[width=18.0cm]{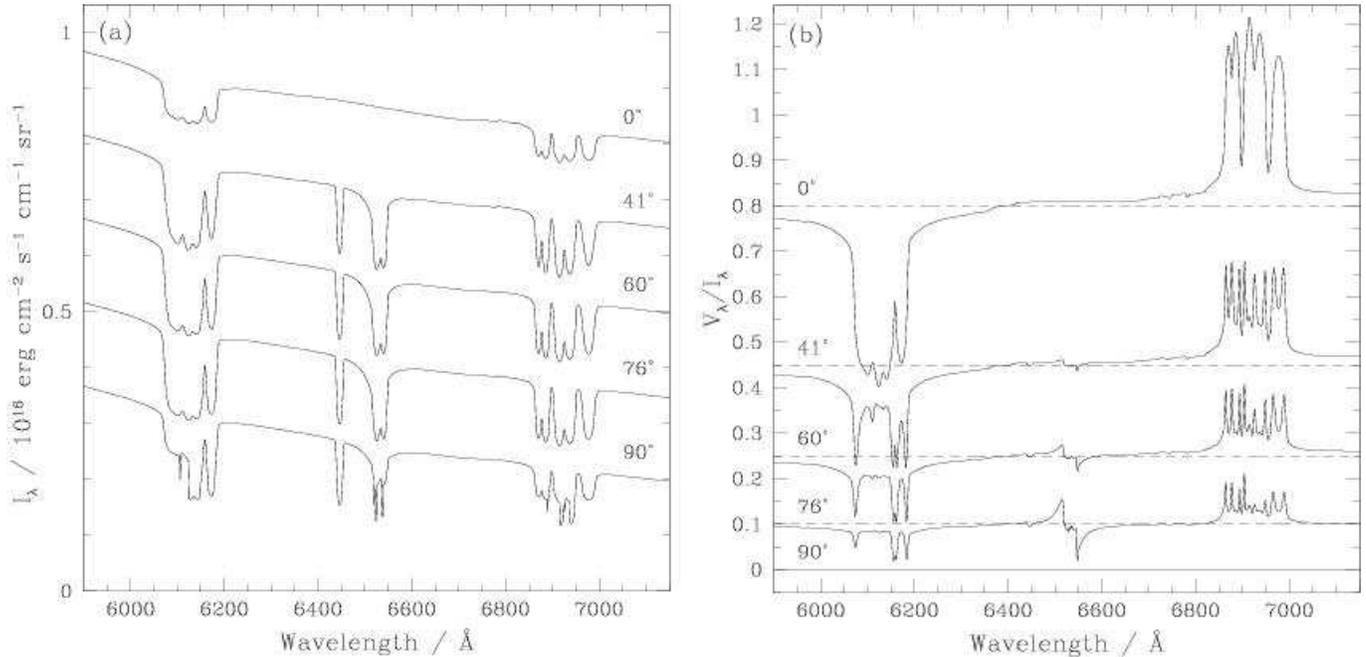}
\caption{Examples of the database spectra for \mbox{$B = 20$\,MG},
\mbox{$T = 15\,000$\,K}, and five angles of $\psi$ equally spaced in
$\cos \psi$ covering the H$\alpha$ $\sigma^-$-, $\pi$-, and
$\sigma^+$-components. \textbf{a)} Intensity, \textbf{b)} degree of
circular polarization. The spectra are shifted vertically by arbitrary
amounts to avoid overlap.}
\label{fig:database_models}
\end{figure*}

We computed a three-dimensional grid of Stokes $I$ and $V$ model
spectra with the effective atmospheric temperature $T$, the magnetic
field strength $B$, and the field direction $\psi$ relative to the
line of sight as the independent variables. We considered $T$ =
8000~K, 9000~K, 10\,000~K, 11\,000~K, 12\,000~K, 13\,000~K, 15\,000~K,
17\,000~K, 20\,000~K, 25\,000~K, 30\,000~K, 40\,000~K, $B = 10$~MG to
200~MG in steps of 1~MG, and $\psi$ = 0$^{\circ}$, 29$^{\circ}$,
41$^{\circ}$, 51$^{\circ}$, 60$^{\circ}$, 68$^{\circ}$, 76$^{\circ}$,
82$^{\circ}$, 90$^{\circ}$, i.e., equidistant in $\cos \psi$. This
yields a database containing \mbox{12 $\times$ 191 $\times$ 9 =
20\,628} model spectra for $I$ and $V$ each. All spectra are
calculated for a surface gravity of \mbox{$\log g = 8$}. Since we do
not include the linear polarization, the field direction is
sufficiently constrained by the total field strength and the
longitudinal field component.  As an example,
Fig.~\ref{fig:database_models} shows a section around H$\alpha$ for a
sample of database spectra with \mbox{$T = 15\,000$\,K}, \mbox{$B =
20$\,MG} and five angles of $\psi$, equally spaced in $\cos \psi$.  A
typical property of these Zeeman spectra is the weak angular
dependence of Stokes $I$, except near 0\degr\ and 90\degr, and the
more pronounced dependence of Stokes $V$. Somewhat simplified, Stokes
$I$ carries much of the information on the distribution of the
absolute value of $\vec{B}$ over the surface of the star, while Stokes
$V$ is needed to derive the distribution of the field directions.

If limb darkening is important the direction cosine $\mu$ of the line
of sight with respect to the vertical direction in the stellar
atmosphere needs to be considered as a further parameter in the
database. Hence, including a wavelength-dependent description of limb
darkening requires an expansion of the number of model spectra in the
database by a factor equal to the number of \mbox{$\mu$-values}
considered. For the present calculations, we use a simple limb
darkening law which is independent of wavelength and avoid this
extension of the database (see Sect.~\ref{sec:ld} below).

\section{Magnetic field geometry}
\label{sec:mfgeometry}

\begin{figure*}[t]
\includegraphics[width=18.0cm]{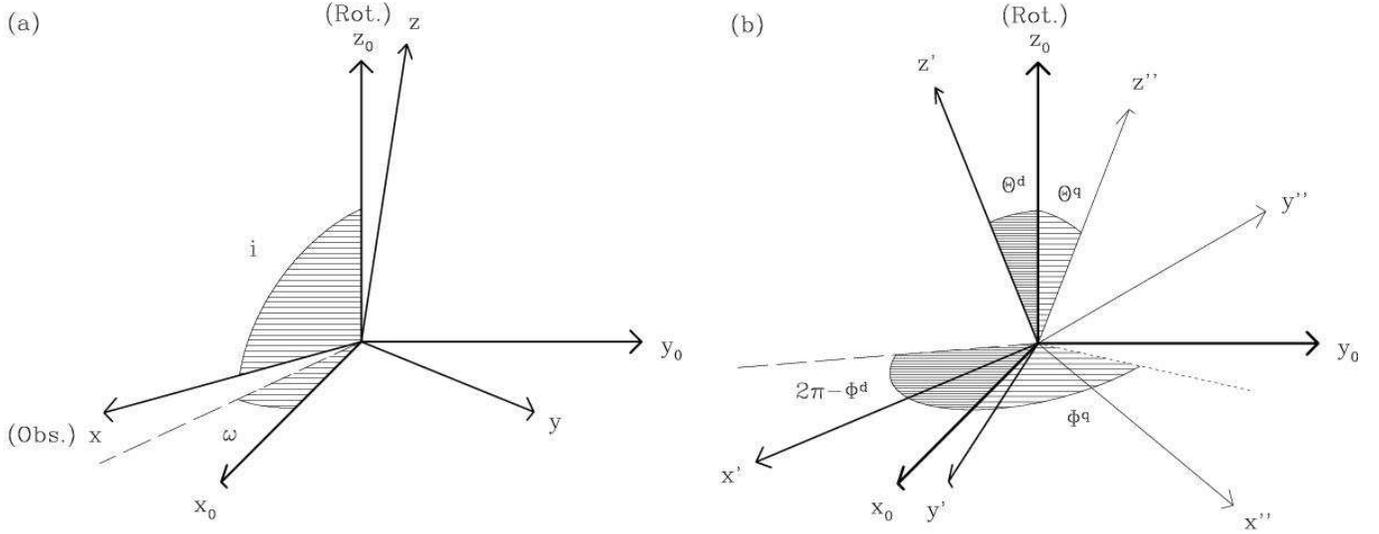} 
\caption{\textbf{a)} Rotational geometry of the MWD models.  The
$x$-axis points towards the observer, the $z_0$-axis marks the
rotation axis. The dashed line marks the intersection of the
$x$-$z$-plane with the $x_0$-$y_0$-plane. $i$ denotes the inclination,
and $\omega$ the rotational phase angle. \textbf{b)} Magnetic geometry
of the MWD models.  The $z'$- and $z''$-axes mark the axes of symmetry
for the dipole and quadrupole components. The lines of intersection of
the $x'$-$z'$-plane and the $x''$-$z''$-plane with the
$x_0$-$y_0$-plane are given by the dashed and the dotted lines,
respectively.}  
\label{fig:geometry}
\end{figure*}

A curl-free field which originates only from sources in the stellar
interior can be described by a multipole expansion of the scalar
magnetic potential, using spherical harmonics with coefficients $l$
and \mbox{$m = 0,\dots,l$}, which describe the zonal and sectoral
periodicity of the field, respectively \citep[see,
e.g.,][]{langel87-1}. The number of free parameters of the field
geometry is \mbox{$l(l+2)$}, i.e. 8~(15) for
\mbox{$l=2$~(3)}. Defining the viewing geometry requires three
additional parameters for the orientation of the rotation axis and the
inclination.

The optimization procedure adopted for the present tests can handle
only a limited number of multipole parameters and becomes inefficient
already when the octupoles \mbox{($l=3$)} are included.  We have
restricted the complexity of the field, therefore, by including only
the two lowest zonal harmonics, commonly referred to as `dipole'
(\mbox{$l=1$}, \mbox{$m=0$}) and `quadrupole' (\mbox{$l=2$},
\mbox{$m=0$}), allowing their axes to be inclined with respect to each
other. We do not consider the \mbox{$m=1$} and \mbox{$m=2$}
quadrupoles, but instead include a common offset of the
dipole-quadrupole combination from the centre of the star. This hybrid
model has ten free parameters: two polar field strengths; two angles
each for the directions of the axes relative to the rotation axis; the
three components of the offset; and the inclination, i.e. the angle
between the rotation axis and the line of sight. An offset from the
centre was included because of its popularity and simplicity (e.g.,
the Earth's magnetic field is approximately that of a shifted
dipole). The chosen field structure deliberately includes some very
similar field geometries described by different sets of parameters: a
combination of aligned dipole and quadrupole can be approximated by a
shift of the dipole. At sufficient signal to noise, the reconstruction
procedure can distinguish between such geometries, a result which is
of interest by itself. While our hybrid model is useful for the tests
performed in this paper, its limited complexity may not suffice for
the interpretation of real, observed spectra.

We consider rotating MWDs viewed at an inclination $i$ with respect to
the rotation axis.  Note that a fraction \mbox{$f = 0.5\,(1
-\sin\,i)$} of the stellar surface is permanently hidden from view and
that phase-resolved Zeeman spectroscopy provides no information on the
field on this hidden fraction of the surface. In order to save
computing time, we restrict ourselves to simultaneously fitting flux
and polarization spectra at four rotational phases, \mbox{$\phi$ =
0.0, 0.25, 0.5, and 0.75}. We avoid a special geometry by choosing
\mbox{$\phi = 0$} not to coincide with the nearest approach of one of
the axes to the line of sight.

Since observational restrictions often prevent taking phase-resolved
data, we also consider the amount of information which can be
retrieved from a single flux and polarization spectrum or even a
single flux spectrum only. In this case, the data provide information
on the magnetic field structure only for one hemisphere of the star.

At any given phase $\phi$, the polarization depends on the components
of the field transverse and parallel to the line of sight. In order to
describe these components, we introduce four Cartesian coordinate
systems (Figs.~\ref{fig:geometry}a and \ref{fig:geometry}b): (i, ii)
systems $\Sigma'$ and $\Sigma''$, in which $z'$ and $z''$ describe the
dipolar and quadrupolar axes of symmetry, respectively; (iii) the
observer's system $\Sigma$, in which the $x$-axis points towards the
observer and the $z$-axis lies in the plane defined by the $x$-axis
and the rotation axis; and (iv) the auxiliary system $\Sigma_0$ with
$z_0$ the direction of the rotation axis which defines the inclination
angle $i$. The rotational phase angle \mbox{$\omega = 2\pi\phi$} is
defined with respect to the direction of the $x_0$-axis which lies in
the $x$-$z$-plane for \mbox{$\omega=0$}.

The components of the surface field $\vec{B}(\vec{r}')$ of the centred
dipole in system $\Sigma'$ are
\begin{align}
(B^\mathrm{d})_{x'} &= 3 B^\mathrm{d}_\mathrm{pol}\,x' z' / (2 {r'}^5)\,,\\
(B^\mathrm{d})_{y'} &= 3 B^\mathrm{d}_\mathrm{pol}\,y' z' / (2 {r'}^5)\,,\\
(B^\mathrm{d})_{z'} &= B^\mathrm{d}_\mathrm{pol} 
                (3 {z'}^2 - {r'}^2) / (2 {r'}^5)\,,
\end{align} 
with $\vec{r}' = (x', y', z'$) and \mbox{$|{\vec{r}'}|^2 = {r'}^2 =
{x'}^2 + {y'}^2 + {z'}^2$}. Cor\-respondingly, the components of the
centred quadrupole in $\Sigma''$ are
\begin{align}
(B^\mathrm{q})_{x''} &= B^\mathrm{q}_\mathrm{pol}\,x'' (5 {z''}^2 - {r''}^2) / 
                (2 {r''}^7)\,,\\
(B^\mathrm{q})_{y''} &= B^\mathrm{q}_\mathrm{pol}\,y'' (5 {z''}^2 - {r''}^2) / 
                (2 {r''}^7)\,,\\
(B^\mathrm{q})_{z''} &= B^\mathrm{q}_\mathrm{pol}\,z''(5 {z''}^2 - 3 {r''}^2)/ 
                (2 {r''}^7)\,,
\end{align}
with $\vec{r}'' = (x'', y'', z'')$ and $|{\vec{r}''}|^2 = {r''}^2 =
{x''}^2 + {y''}^2 + {z''}^2$. $\Sigma'$ and $\Sigma''$ are tilted with
respect to the rotation axis $z_0$ by angles \thed\ and \theq,
respectively. The azimuth angles of the tilt in $\Sigma_0$ are \phid\
and \phiq\ at phase \mbox{$\phi = 0$}. We apply the appropriate
rotations and the translation to transform the fields into the
observer's system $\Sigma$, add the dipole and quadrupole components,
and obtain $\vec{B}(\vec{r})$ for each surface element, with
\mbox{$B_x = B_\mathrm{l} = B \cos\psi$} the longitudinal component of
the field.

The angle $\delta$ between the dipole and quadrupole axes and the
angles $\eta^\mathrm{d}$ and $\eta^\mathrm{q}$ between the line of
sight at phase $\phi$ and the dipole and quadrupole axis,
respectively, are given by
\begin{gather}
\cos \delta = \cos \Theta^\mathrm{d} \cos \Theta^\mathrm{q} +
             \sin \Theta^\mathrm{d} \sin \Theta^\mathrm{q}
              \cos (\Phi^\mathrm{q} - \Phi^\mathrm{d})\,,\\
\cos \eta^\mathrm{d} = \cos i\,\cos \Theta^\mathrm{d} +
             \sin i\,\sin \Theta^\mathrm{d}
              \cos (2\pi\phi + \Phi^\mathrm{d})\,,\\
\cos \eta^\mathrm{q} = \cos i\,\cos \Theta^\mathrm{q} +
              \sin i\,\sin \Theta^\mathrm{q}
              \cos (2\pi\phi + \Phi^\mathrm{q})\,.
\end{gather}

Our magnetic geometries were selected for the purpose of providing
sufficient complexity for an effective test of our reconstruction
routine. The offset \roff\ from the centre is a simple means of
producing a substantial amount of azimuthal asymmetry if \roff\ is
perpendicular to the dipole axis, while \roff\ parallel to the dipole
axis allows to test the ability of the routine to distinguish between
aligned centred dipole-quadrupole combinations and a shifted dipole.

\section{Input spectra for the reconstruction procedure}
\label{sec:input_spectra}

\subsection{Integration of the database models}
\label{sec:integration}

\begin{figure}[t]
\includegraphics[width=8.8cm]{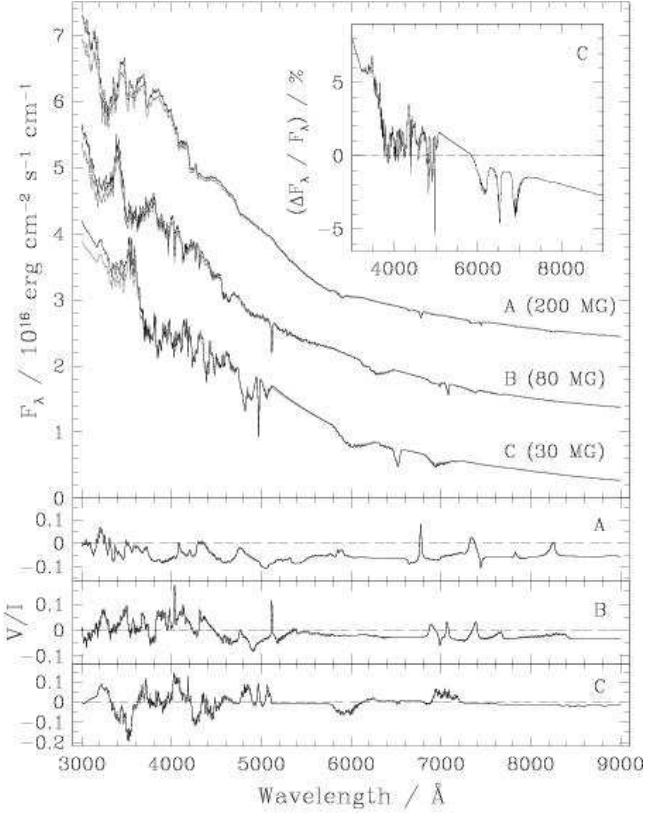}
\caption{Spectral flux \textit{(top)} and circular polarization
\textit{(bottom)} for centred dipoles viewed pole-on with polar field
strengths of (A) 200\,MG, (B) 80\,MG, and (C) 30\,MG for \mbox{$T$ =
15\,000\,K}. The library-based spectra computed using a database of
four $\mu$-values (grey) are compared to those for $\mu = 1$,
corrected with a mean limb darkening law (black). Spectra (A) and (B)
have been shifted upwards to avoid overlap (1.1 flux units each). The
insert shows the relative flux differences for the 30\,MG case (C).}
\label{fig:cmpsc}
\end{figure}

We divide the spherical star into surface elements defined by equal
steps in latitude and longitude. For given distributions of the
magnetic field vector $\vec{B}$ and the effective temperature $T$ over
the surface, let $\alpha$ be the running index of the surface elements
which are visible at a given rotational phase \mbox{$0 \geq \phi \geq
1$} and which have sizes $A_\alpha$, central field strengths
$B_\alpha$, field directions $\psi_\alpha$, and direction cosines
$\mu_\alpha$. The Stokes parameters $\left< I_\lambda \right>$ and
$\left< V_\lambda \right>$ are then computed as weighted sums of the
individual contributions corrected for limb darkening by a factor
$f^{\mathrm{LD}}_{\alpha}$ (discussed below)
\begin{equation}
\begin{pmatrix} \left< I_\lambda \right> \\ \left< V_\lambda \right> 
        \end{pmatrix} (\phi) = \sum_{\alpha(\phi)} A_{\alpha}
        \mu_{\alpha} f^{\mathrm{LD}}_{\alpha} \begin{pmatrix}
        I_{\lambda,\alpha} \\ V_{\lambda,\alpha}
\end{pmatrix}. 
\end{equation}
We represent the wavelength-dependent contributions
$I_{\lambda,\alpha}$ and $V_{\lambda,\alpha}$ from surface element
$\alpha$ by appropriate interpolation in the database grids of the
parameters $T$, $B$, and $\psi$.
For $T$ and $\psi$, a bilinear interpolation suffices. For the field 
strength, we consider all
spectra representative of the $B$-variation over the finite surface
element. We found that a number of 900 surface elements per hemisphere
is a good compromise between CPU time and needed accuracy.  This
number is sufficient to avoid spectral structure caused by the finite
element size.

Our code can account for temperature variations over the surface of
the white dwarf, but in this paper we consider only stars with uniform
surface temperatures.

\begin{figure*}[t]
\includegraphics[width=18.0cm]{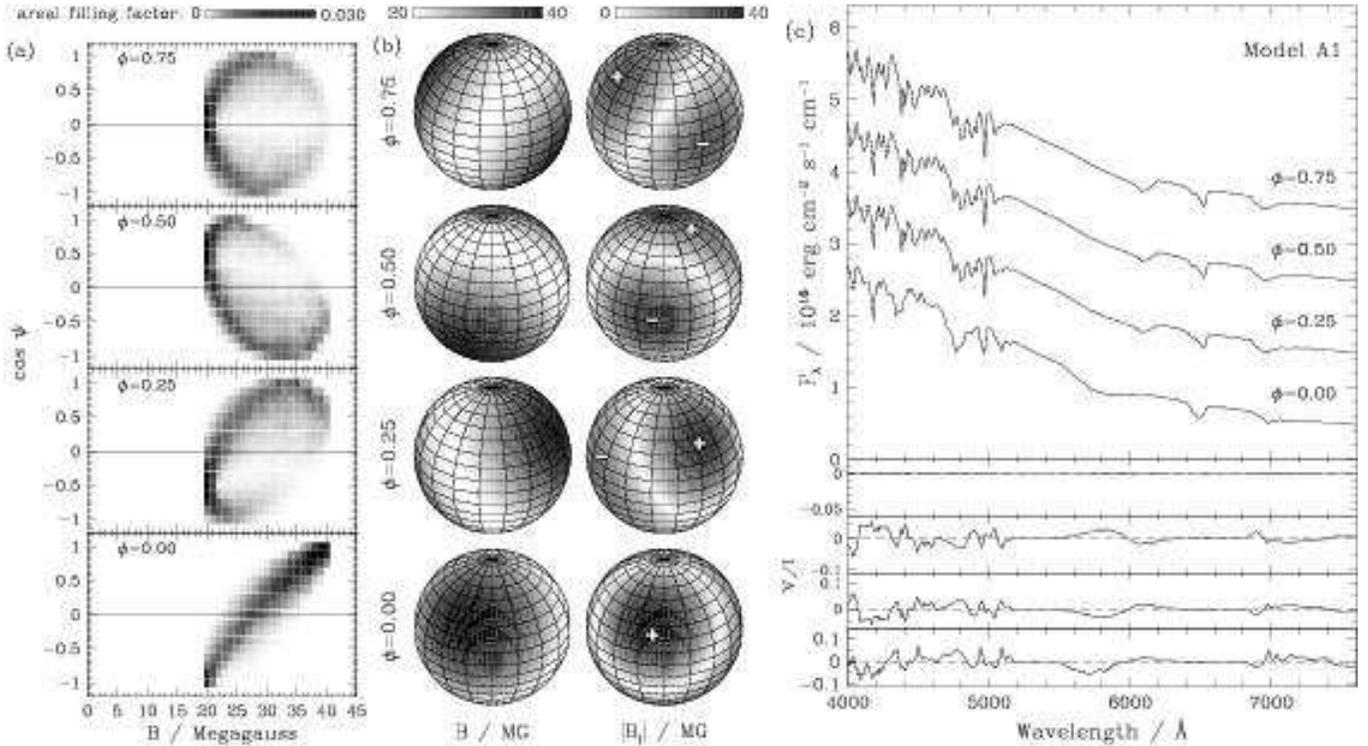} 
\caption{Model (A1), centred dipole viewed at $i = 60^\circ$:
\textbf{a)} \bpd, \textbf{b)} field strength and longitudinal
component, \textbf{c)} flux and polarization spectra. The $+$ and $-$
symbols indicate the sign of the longitudinal component of the
magnetic field. For clarity, the flux spectra at \mbox{$\phi$ = 0.25,
0.5, and 0.75} have been shifted upwards by one flux unit each.}
\label{fig:A1}
\end{figure*}

\subsection{Relation to the observed flux}

The flux observed from a star of radius $R$ at distance $d$ is
\begin{equation}
f_{\lambda} = F_{\lambda}\,R^2/d^2
\end{equation}
where \mbox{$F_{\lambda} = \pi\left< I_\lambda \right>$} is a function
of $T$, for a given magnetic field distribution. The interpretation of
observed Zeeman spectra in terms of $\left< I_\lambda \right>$ and
$\left< V_\lambda \right>$ involves $T$ and $R/d$ as fit parameters.

For the present tests, $T$ and $R/d$ are considered as fixed
parameters and the quantities fitted by variation of the field
parameters are $F_{\lambda}$ and \mbox{$V/I = \left< V_\lambda \right>
/ \left< I_\lambda \right>$}.

\subsection{Limb darkening}
\label{sec:ld}

We have compared (i) the full radiative transfer for each surface
element which accounts for the $\mu$-dependence and the variation of
$B$ across the element already in the atomic data, (ii) an
interpolation between the spectra for discretized $\mu$ and $B$, and
(iii) the application of a wavelength-independent linear limb
darkening law replacing the interpolation in $\mu$.  Method (ii) uses
spectra calculated for \mbox{$\mu$ = 0.1, 0.4, 0.7, and 1.0}. Method
(iii) employs a linear law with coefficients which are valid for the
visible wavelength range,
\begin{equation}
f^{\mathrm{LD}}_{\alpha} = I_{\lambda}(\mu) / I_{\lambda,\mu = 1} 
= 0.70 + 0.30 \mu\,.
\label{eq:ld}
\end{equation}
Test calculations for the three approaches (i) to (iii) were
performed for centred dipoles viewed pole-on with polar field
strengths of (A) 200\,MG, (B) 80\,MG, and (C) 30\,MG.  We found the
differences between (i) and (ii) to be minute.  Case (iii) differs by
a wavelength-dependent factor which arises from the neglect of any
wavelength dependence in the limb darkening approximation.
Figure~\ref{fig:cmpsc} shows the results for cases (ii) and (iii) at an
effective temperature of 15\,000\,K.  For all three field strengths,
the spectra computed for case (iii) deviate by at most 5\% from those
of cases (i) and (ii).  The insert shows that the absorption lines are
about 2\% deeper than for the correct treatment.

The simple limb darkening law of Eq.~\ref{eq:ld} is entirely
acceptable for the present tests which interpret synthetic spectra
with spectra of the same origin. The above comparison suggests,
moreover, that a wavelength-independent linear limb darkening law may
also be acceptable for the interpretation of observed optical Zeeman
spectra of white dwarfs.

\section{Reconstruction of the field geometry}
\label{sec:reconstruction}

In this Section, we describe a variety of magnetic field and viewing
geometries and test the ability of our code to reconstruct their
parameters from flux and circular polarization spectra at \mbox{$\phi$
= 0.0, 0.25, 0.5, and 0.75}. All calculations were performed for an
effective temperature of 15\,000\,K.  In order to simulate real data,
noise was added to the input spectra at the four phases as described
below. Because of the added noise, the reconstructed field is not
necessarily identical to the input field. For the present tests, the
wavelength range was restricted to \mbox{$4000 \leq \lambda \leq
7600$\,\AA}, which contains the most important Balmer line components,
and all spectra were rebinned into 10 \AA\ bins, yielding 361 data
pixels per spectrum, a total of 1444 pixels in the combined flux
spectra at four rotational phases, and another 1444 pixels in the
polarization spectra.

\subsection{Magnetic field models}
\label{sec:reference_spec}

\begin{figure*}[t]
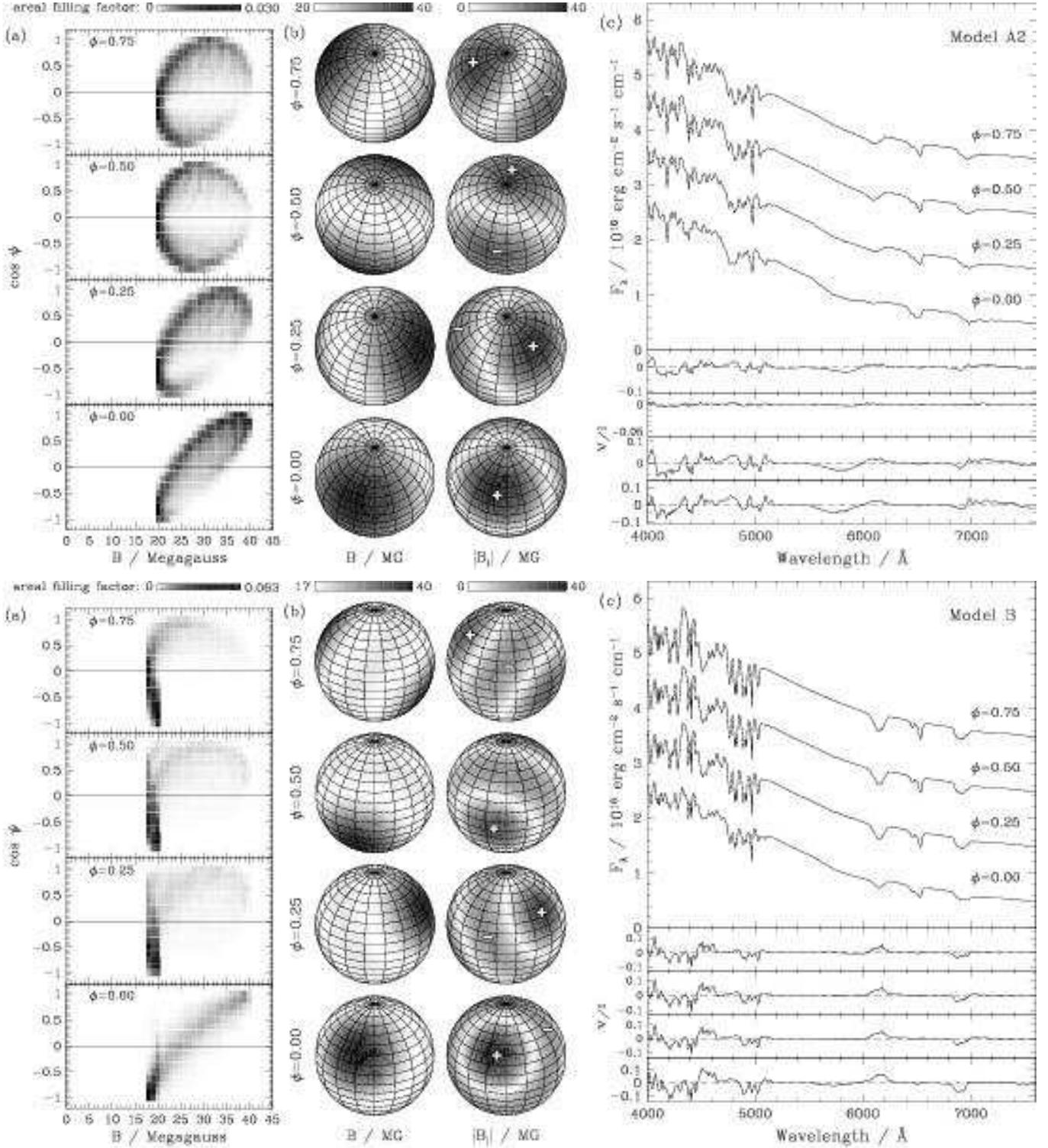

\includegraphics[width=18.0cm]{h3480.f5t} 

\vspace*{1ex}
\includegraphics[width=18.0cm]{h3480.f5b} 
\caption{\emph{Top:} Model (A2), centred dipole viewed at $i =
30^\circ$: \textbf{a)} \bpd, \textbf{b)} field strength and
longitudinal component, \textbf{c)} flux and polarization spectra.
\emph{Bottom:} Model (B), pure quadrupole viewed at $i = 60^\circ$:
\textbf{a)} \bpd, \textbf{b)} field strength and longitudinal
component, \textbf{c)} flux and polarization spectra. See
Fig.~\ref{fig:A1} for further explanation.}
\label{fig:A2+B}
\end{figure*}

\begin{figure*}[t]
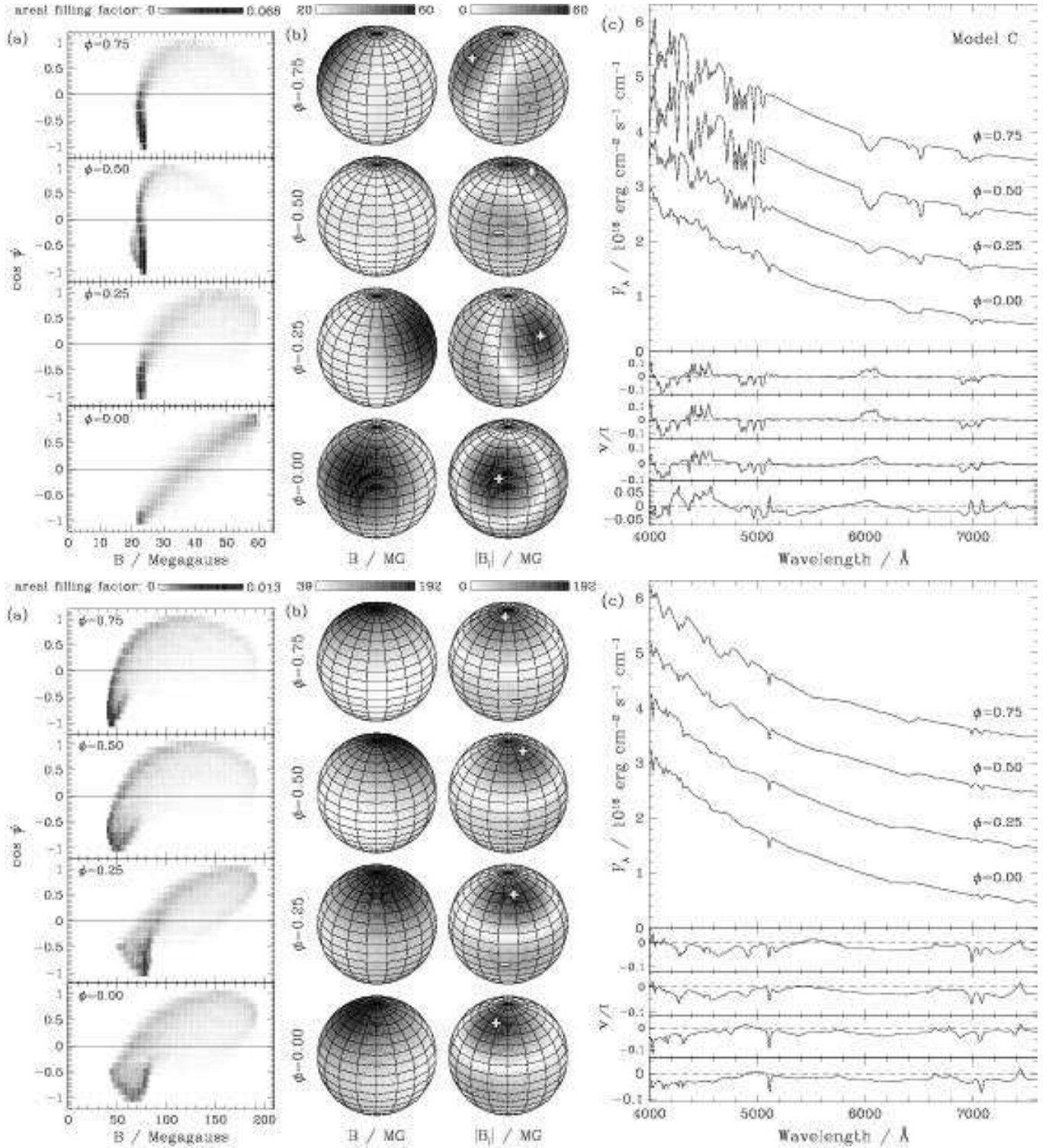
 
\includegraphics[width=18.0cm]{h3480.f6t} 

\vspace*{1ex}
\includegraphics[width=18.0cm]{h3480.f6b} 
\caption{\emph{Top:} Model (C), aligned dipole and quadrupole viewed
at $i = 60^\circ$: \textbf{a)} \bpd, \textbf{b)} field strength and
longitudinal component, \textbf{c)} flux and polarization spectra.
\emph{Bottom:} Model (D), shifted high-field dipole viewed at $i =
60^\circ$: \textbf{a)} \bpd, \textbf{b)} field strength and
longitudinal component, \textbf{c)} flux and polarization spectra. See
Fig.~\ref{fig:A1} for further explanation.}
\label{fig:C+D}
\end{figure*}

\begin{figure*}[t]
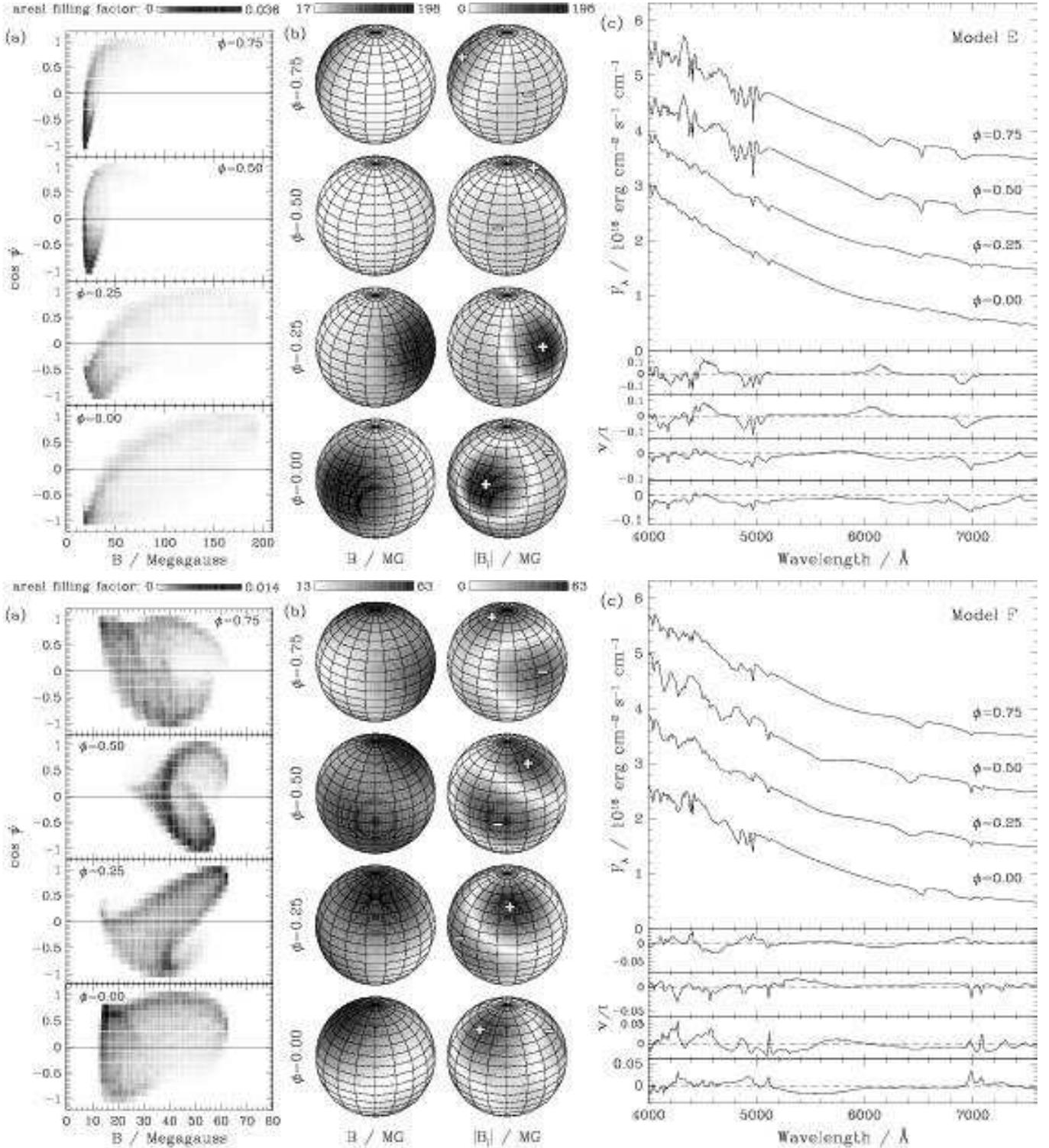

\includegraphics[width=18.0cm]{h3480.f7t} 

\vspace*{1ex}
\includegraphics[width=18.0cm]{h3480.f7b} 
\caption{\emph{Top:} Model (E), shifted dipole viewed at $i =
60^\circ$: \textbf{a)} \bpd, \textbf{b)} field strength and
longitudinal component, \textbf{c)} flux and polarization spectra. 
\emph{Bottom:} Model (F), non-aligned dipole-quadrupole combination viewed
at $i = 60^\circ$: \textbf{a)} \bpd, \textbf{b)} field strength and
longitudinal component, \textbf{c)} flux and polarization spectra. See
Fig.~\ref{fig:A1} for further explanation.}
\label{fig:E+F}
\end{figure*}

We define seven different magnetic field and viewing geometries
against which we test our reconstruction code. The field
configurations (A) to (F) are characterized by an increasing level of
complexity.  The geometrical and spectral properties of the models are
summarized in Figs.~\ref{fig:A1}--\ref{fig:E+F}. In each case, the
centre panel shows the distributions of the total field strength $B$
and of the absolute value of the longitudinal component $B_\mathrm{l}$
over the visible hemisphere at the four selected phases. The $+$ and
$-$ symbols indicate the sign of the longitudinal component. The range
of field strengths realized over the visible part of the surface of
the white dwarf is given by the top grey bar. The left-hand panel
shows the \bpd, a greyscale plot of the frequency distribution of the
magnetic field strength $B$ and the direction cosine $\cos \psi$.  The
fractional contribution of each single database spectrum to the
integrated spectrum is represented by the greyscale value of the
corresponding pixel in the plot. This presentation includes the
effects of pixel area, foreshortening, and limb darkening. The sum of
all filling factors would be unity if limb darkening were neglected,
but falls below unity with limb darkening included. The {\bpd}s are
equivalent to the ZEBRA plots of \citet{donatietal94-1}, except for
the effect of limb darkening which was not included by these
authors. The diagrams illustrate the change in the weighting of the
two main database parameters, $B$ and $\psi$, as the star rotates. The
right-hand panel shows the resulting integrated flux and circular
polarization spectra at the four rotational phases.

{\itshape Model (A1), centred dipole viewed at $i=60\degr$:} The polar
field strength is \mbox{\bdip\ = 40\,MG} and the axis points towards
(\thed, \phid) = (60\degr, 340\degr). This oblique rotator model
stands for a simple low-field geometry. The hidden fraction is only
7\% of the white dwarf surface. The flux spectra in Fig.~\ref{fig:A1}c
are quite similar at $\phi = 0.25, 0.5$, and 0.75, but the circular
polarization spectra are not. The \bpd\ looks different at \mbox{$\phi
= 0$} and so does the flux spectrum. These differences suggest that
full phase coverage is essential for a successful recovery of the
field geometry.

{\itshape Model (A2), centred dipole viewed at $i=30\degr$:} The
hidden fraction of the surface is now 25\%. Otherwise, the properties
of the model (Fig.~\ref{fig:A2+B}, top) are similar to (A1). For the
centred dipoles of models (A1) and (A2), the circular polarization
vanishes at a phase $\phi_0$, where the dipole axis is oriented
perpendicular to the line of sight.

{\itshape Model (B), centred quadrupole viewed at $i=60\degr$:} The
polar field strength is \mbox{\bqua\ = 40\,MG} and the axis points
towards (\theq, \phiq) = (60\degr, 340\degr). Figure~\ref{fig:A2+B}c
(bottom) shows that there is little rotational variation. The flux
spectra and the polarization vary little for $\phi = 0.25, 0.5$, and
0.75, but differ at \mbox{$\phi = 0$}.

{\itshape Model (C), aligned centred dipole and quadrupole viewed at
$i=60\degr$:} A quadrupole of \mbox{\bqua\ = 20\,MG} is added to a
dipole of \mbox{\bdip\ = 40\,MG} with (\thed, \phid) = (\theq, \phiq)
= (60\degr, 340\degr). The asymmetry introduced into the field
geometry causes larger rotational variations in flux and polarization
than for the pure dipole or quadrupole (Fig.~\ref{fig:C+D}c, top).

{\itshape Model (D), shifted high-field dipole viewed at $i=60\degr$:}
The polar field strength is \mbox{\bdip = 110\,MG}, offset in all
three coordinates by (\xoff, \yoff, \zoff) = (0.05, $-$0.10,
0.15). The shift along the dipole axis increases the maximum field,
and the sideways shift decreases the minimum field to the effect that
$B$ ranges from 39 to 192\,MG. That is, $B$ varies by a factor of five
compared to a factor of two for the centred dipole.  The high field
causes the flux spectra to show substantially less structure than in
the previous models, suggesting that a higher signal-to-noise ratio is
needed for reconstruction (Fig.~\ref{fig:C+D}c, bottom). There is
substantial variation in the circular polarization over the rotational
period, however, which helps in the reconstruction.

{\itshape Model (E), shifted dipole viewed at $i=60\degr$:} This is an
extremely off-centred dipole with \mbox{\bdip\ = 58\,MG} and (\xoff,
\yoff, \zoff) = (0.15, $-$0.10, 0.30) which displays a variation of
$B$ over the surface by nearly a factor of 12. For one half of the
rotational period, the high-field pole is in view, over the other half
the field distribution is concentrated at low field
strengths. Effectively, this represents a star with a low field of
around 20\,MG over most of the star and a spot in which the field
rises to 198\,MG. The circular polarization displays pronounced
rotational structure (Fig.~\ref{fig:E+F}c, top).

{\itshape Model (F), non-aligned dipole-quadrupole combination viewed at
$i=60\degr$:} This is the most complex field model featuring the
superposition of a non-aligned dipole and quadrupole with equal polar
field strengths of 40\,MG.  The polar directions, (\thed, \phid) =
(60\degr, 340\degr) and (\theq, \phiq) = (30\degr, 250\degr), are
separated by 64\degr. The field geometry features two high-field
spots, an upper positive and a lower negative one, which are dominated
by the quadrupole and the dipole, respectively, and are separated by
$\sim$90\degr\ (Fig.~\ref{fig:E+F}, bottom).

\begin{figure}[t]
\includegraphics[width=8.8cm]{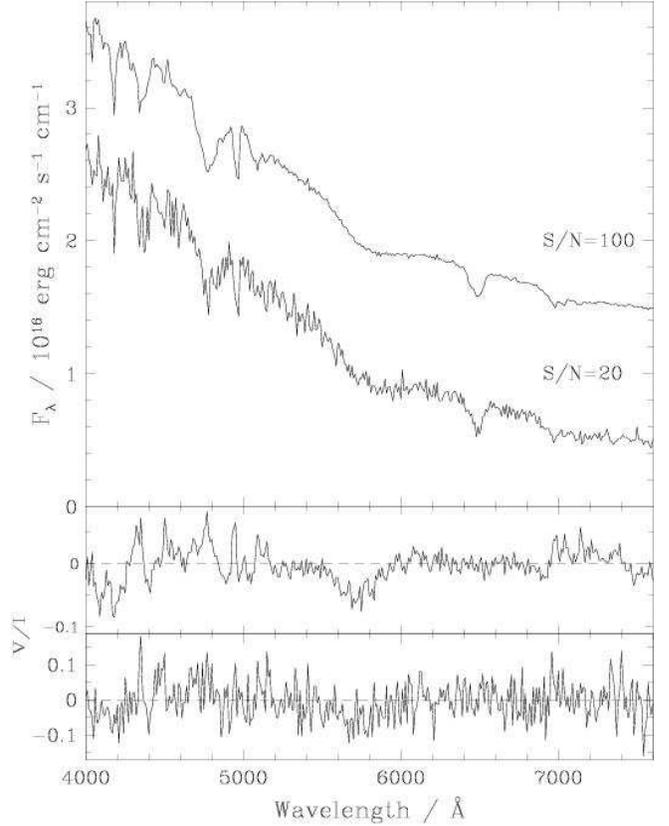}
\caption{Examples of simulated spectra used as input to the
reconstruction procedure. The \mbox{$\phi = 0$} spectrum of Model (A1) is
shown with a signal-to-noise ratio \mbox{$S/N=20$} and 100 (shifted upwards
by one flux unit). \emph{Top:} Flux spectra, \emph{bottom:} circular
polarization spectra.}
\label{fig:noisified}
\end{figure}

An overview of the models (A)--(F) is given in Table~\ref{tab:input}.
To illustrate the effects of noise in the spectra used in the
reconstructions, we show in Fig.~\ref{fig:noisified} the flux and
polarization spectra of Model (A1) for \mbox{$\phi=0$} at noise levels
of \mbox{$S/N=100$} and \mbox{$S/N=20$}.

\subsection{Optimization algorithm}

Our spectral synthesis method is sufficiently fast to allow the use of
hierarchical search strategies in the parameter space. We utilize the
optimization routine \texttt{evoC} \citep{trint+utecht94-1} that
implements an evolutionary strategy algorithm \citep{rechenberg94-1},
and has proven useful already in other astrophysical
contexts~\citep{gaensicke+beuermann96-1, gaensickeetal98-1,
kubeetal00-1}. The task is to find a set \mbox{$\vec{a} = (a_1,
\hdots, a_M)$} of $M$ free parameters that minimizes the classic
penalty function
\begin{equation}
\chi^2_{\mathrm{red}} (\vec{a}) = \frac{1}{N-M} \sum_{j=1}^N 
\frac{(f_j - s_j(\vec{a}))^2} {{\sigma_j}^2}
\label{eq:chisq}
\end{equation}
given the input data pixels $f_j$, the model data pixels $s_j$, and
the standard deviations $\sigma_j$.  Good fits require \mbox{\chisqred
$\approx$~1}. We have applied Gaussian noise to the input spectra to
yield signal-to-noise ratios, corresponding to relative standard
deviations in $F_\lambda$ and absolute standard deviations in $V/I$,
of 0.01 and 0.05, respectively.  For fits to fluxes only, $j$ runs up
to \mbox{$N=1444$}, and, for fits to both flux and polarization, up to
\mbox{$N=2888$}.

For each field model and each reconstruction with a certain set of
free parameters, the \texttt{evoC} optimization process is run
repeatedly, typically 6--20 times, starting each run with different,
randomly chosen parameter values. Not all runs end up in the global
minimum.  A misguided run may be caught in a local minimum
corresponding to an incorrect field configuration, which nevertheless
has a Zeeman spectrum similar to the input one. We define a success
rate of the optimization as the fractional number of runs which reach
a best-fit \mbox{\chisqred\ $<$ 2.0} for \mbox{$S/N=100$}, and
\mbox{\chisqred\ $<$ 1.1} for \mbox{$S/N=20$} (corresponding virtually
always to the global minimum).  As a last finish, we employ a downhill
simplex algorithm \citep{nelder+mead65-1,pressetal92-1} on the run
with the best \chisqred, which sometimes improves on the \texttt{evoC}
solution.

\begin{figure*}[t]
\includegraphics[width=18.0cm]{h3480.f9}
\caption{\textit{Left panel:} Contour plot of the \chisq-landscape for
the spectral flux in the \bqua,\theq-plane. The input configuration is
given by \mbox{\bdip\ = 40\,MG}, \mbox{\bqua\ = 20\,MG}, and
\mbox{\thed\ = \theq\ = 0\degr}. Darker shading corresponds to smaller
values of \chisqred. \textit{Centre panel:} {\bpd}s of the field
configurations corresponding to the local \textit{(top)} and the
global \textit{(bottom)} minimum. \textit{Right panel:} Flux and
circular polarization spectra corresponding to the the global minimum
(lower curves) and the local minimum in the upper left of the left
panel (upper curves, shifted upwards by 0.5 units in flux and 0.1
units in polarization).}
\label{fig:chisq}
\end{figure*}

In order to illustrate the problem associated with local minima in the
\chisq-landscape, we present in Fig.~\ref{fig:chisq} a simple example
of different field geometries which yield similar Zeeman spectra. The
input geometry is the sum of a (non-rotating) dipole with \mbox{\bdip\
= 40\,MG} and an aligned quadrupole with \mbox{\bqua\ = 20\,MG},
viewed at \mbox{$i=60\degr$} (with \thed\ = \theq\ = $\omega =
0$\degr, Fig.~\ref{fig:geometry}). The right-hand panel of
Fig.~\ref{fig:chisq} shows the corresponding flux and polarization
spectra (lower curves). We add Gaussian noise of \mbox{$S/N=100$} and
compute flux and flux+polarization spectra covering a range of
quadrupole parameters, with the dipole parameters and the inclination
kept fixed. The quadrupole is allowed to vary in strength and
orientation with \theq\ free at \mbox{\phiq\ = 90\degr}. For this
choice of parameters, the quadrupole is perpendicular to the dipole
and to the line of sight for \mbox{\theq\ = 90\degr}.  The left-hand
panel of Fig.~\ref{fig:chisq} shows a contour plot of the
\chisq-landscape for the spectral flux in the
\bqua,\theq-plane. Besides the global minimum at the parameter values
of the input configuration (\mbox{\bqua\ = 20\,MG}, \mbox{\theq\ =
0\degr}), a second pronounced minimum appears at \mbox{\bqua\ =
$-$17\,MG} and \mbox{\theq\ = 90\degr}, with the minus sign indicating
a reversed polarity of the quadrupole. The flux and circular
polarization spectra for this minimum are also shown in the right-hand
panel of Fig.~\ref{fig:chisq}. At moderate noise levels, the flux and
polarization spectra of these two diverse field geometries become
indistinguishable and it is not surprising that the local minimum (in
the upper left corner of the contour plot) persists if flux and
circular polarization are considered together.  The shallow local
minimum at \mbox{\bqua\ $\simeq 33$\,MG}, \mbox{\theq\ $\simeq
80$\degr}, on the other hand, disappears if $V/I$ is included in the
computation of \chisqred.  In the centre panels of
Fig.~\ref{fig:chisq}, the {\bpd}s for both configurations are shown.
Both distributions are sufficiently similar if projected either on the
$B$-axis or on the $\cos \psi$-axis to explain why the spectra are
similar, but not identical.

Finally, we note that fitting the remaining parameters of the field
model (like \bdip) instead of keeping them fixed would cause the local
minima to become even more pronounced. Increased noise also deepens
the local minima relative to the global one.  A local \chisq-minimum
is responsible for an incorrect, although not entirely dissimilar,
reconstruction of Model~F discussed below.

\subsection{Reconstruction fits}

\subsubsection{General characteristics of the solutions}

Depending on the complexity of the input field, we consider
reconstructions which differ in the numbers of free parameters,
ranging from the full set of ten down to seven (with the quadrupole
component or the offset neglected).  Some redundancy is allowed
because a dipole offset along its axis can also be modelled, to first
order, by an aligned centred dipole-quadrupole combination.  With data
of sufficient $S/N$, the reconstruction procedure can recognize such
subtle differences.

As a general feature, the reconstructed global field is of relevance
only for that part of the stellar surface which is visible during the
observation (or covered by the synthetic input in this paper).  This
underlines the importance of phase-resolved observations which allow
the determination of the inclination and, thereby, to estimate the
occulted fraction of the surface.

Spectrophotometry of high $S/N$ is obtained more easily than
spectropolarimetry of the same quality. The observer has to decide,
therefore, whether a given amount of observing time is better spent on
high-quality spectrophotometry or on circular spectropolarimetry of
lower quality. In order to address such questions, we reconstructed
all field geometries, using the spectral flux \emph{and} the
polarization, and using the spectral flux only.  The flux-only
reconstructions are successful in several cases, but the deviations
from the input geometry tend to be larger, and an increased number of
non-convergent runs suggest a less well-behaved \chisq-landscape. We
find that the circular polarization is not needed in simple cases,
while its inclusion is extremely useful for the reconstruction of more
complex fields.

\subsubsection{Results for individual field geometries}
\label{sec:individual_results}

In this Section, we present the results for the reconstructions of the
input models (A) to (F), using the spectra at four rotational
phases. All results are listed in Table~1.  The column denoted `flag'
indicates whether the fit is to flux \emph{and} polarization (fp) or
to the flux only (f). The last column illustrates the convergence
properties in the multidimensional parameter space, referred to as
success rate above (number of successful runs vs. total number of
runs).

\begin{figure}[t]
\includegraphics[width=8.8cm]{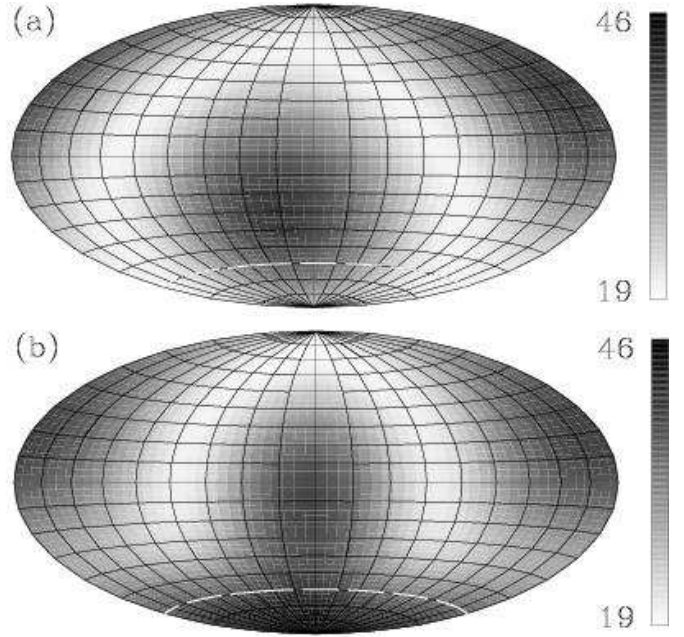}
\caption{Aitoff maps of the magnetic field strength $B$ (in MG) for
Model (A1).  The input \textit{(top)} and the high-noise flux-only
reconstruction of Table~\ref{tab:input}, Model (A1), line~10
\textit{(bottom)} are shown. The axis of the maps represents the
rotation axis. The centre of the maps marks the $-x_0$ direction (see
Fig.~\ref{fig:geometry}b). The region within 30\degr\ from the lower pole is
permanently hidden (dashed white line).}
\label{fig:reconstruction_A1}
\end{figure}

{\itshape Model (A1), centred dipole viewed at $i=60\degr$:} The results in
lines~1--4 assume \mbox{\bqua\ =~0}, those in lines~5--8 zero
offset. All reconstructions are successful and reproduce the dipole
field strength within 0.1\,MG and the magnetic axis and the
inclination with rms deviations of 5\degr and 8\degr,
respectively. Not surprisingly, the accuracy of the reconstruction
benefits from a high $S/N$, but is acceptable even for flux-only fits
and a low $S/N$ ratio. Note that errors in \phid\ are irrelevant as
long as \thed\ matches closely. The same holds for \phiq\ and \theq\
as long as \bqua\ is close to zero. If all parameters are included in
the fit (lines~9 and 10), a quadrupole component usually appears which
is largely compensated for by a shift in the dipole (plus quadrupole)
to the effect that the net field is dipole-like again. The low-noise
flux-and-polarization fit of line~9 is quite acceptable, while the
high-noise flux-only fit of line~10 produces larger misfits in $i$ and
in the field geometry. Even the latter provides an acceptable
reconstruction over the visible part of the surface, but deviates
strongly from the input in the permanently occulted part. This result
is due to the inclusion of a higher multipole component than present
in the input.  Figure~\ref{fig:reconstruction_A1} demonstrates this
result.

{\itshape Model (A2), centred dipole viewed at $i=30\degr$:} The
reconstructions are of a quality similar to that of model (A1). The
permanently occulted fraction of the stellar surface, for which the
reconstruction remains undefined, is now 25\%.

{\itshape Models (B, C), pure quadrupole and aligned dipole-quadrupole
viewed at $i=60\degr$:} The four reconstructions each use the same
set of parameters as lines~5--8 of model (A1). The lack of a dipole
component in (B) and the relative strength of the dipole and
quadrupole components in (C) are recognized even in the flux-only and
high-noise fits.

{\itshape Model (D), shifted high-field dipole viewed at $i=60\degr$:} The
dipole is aligned with the rotation axis. Hence, the rotational
modulation is caused by the off-centre shift. In spite of the weaker
Zeeman structures (Fig.~\ref{fig:C+D}c, bottom), the configuration is
recovered correctly if the search assumes a shifted dipole
(lines~1--4).  The absence of a quadrupole component is recognized in
the low-noise flux-and-polarization fits (line~5), but less so in the
high-noise flux-only fit (line~6). Interestingly, the offset is
recognized correctly in both cases.

{\itshape Model (E), shifted dipole viewed at $i=60\degr$:} This is the
model which features one high-field spot. If modelled as a shifted
dipole, the parameters are recovered with high accuracy, even when the
$S/N$ is low and the polarization is disregarded (lines~1--4).
Allowing for a quadrupole component leads to the usual compensatory
effects (lines~5 and 6). The reconstruction is acceptable over the
visible surface, and deviates only slightly from the input field in
the occulted part.

\begin{table*}[h]
\caption{Reconstructed magnetic parameters for the configurations
(A)--(F). Each model is introduced by a boldface line which lists the
input parameters. The subsequent numbered lines represent the
individual reconstructions. In the `flag' column, `fp' denotes
simultaneous fits to flux and polarization spectra, `f' fits to flux
spectra only. The last column indicates the success rate of the
convergence of the multidimensional parameter search. Note that for
(B) and (C), \theq\ = \thed\ and \phiq\ = \phid. }
\label{tab:input}
\begin{tabular}{crrrrrrrrrrrrcr} 
\hline \noalign{\smallskip}

Model & $i$ & \bdip & \thed & \phid & \bqua & \theq & \phiq & \xoff &
\yoff & \zoff & \chisqred & $S/N$ & flag & conv.\\

 & (\degr) & (MG) & (\degr) & (\degr) & (MG) & (\degr) & (\degr) &
(\rwd) & (\rwd) & (\rwd) & & & & \\ \noalign{\smallskip} \hline
\noalign{\smallskip}


\multicolumn{15}{l}{\itshape (A1), centred dipole viewed at
$i=60\degr$:}\\ & \textbf{60.0} & \textbf{40.0} & \textbf{60.0} &
\textbf{340.0} & \textbf{0.0} & \textbf{0.0} & \textbf{0.0} &
\textbf{0.0} & \textbf{0.0} & \textbf{0.0} & \multicolumn{1}{c}{--} &
\multicolumn{1}{c}{--} & \multicolumn{1}{c}{--} &
\multicolumn{1}{c}{--}\\

1 & 54.4 & 40.0 & 64.2 & 339.0 & \multicolumn{1}{c}{--} &
\multicolumn{1}{c}{--} & \multicolumn{1}{c}{--} & $-$0.001 & 0.000 &
0.000 & 1.07 & 100 & fp & 2/10\\ 2 & 47.0 & 40.2 & 62.8 & 350.6 &
\multicolumn{1}{c}{--} & \multicolumn{1}{c}{--} &
\multicolumn{1}{c}{--} & 0.000 & $-$0.001 & $-$0.003 & 1.01 & 20 & fp
& 2/10\\

3 & 55.3 & 40.3 & 64.1 & 339.6 & \multicolumn{1}{c}{--} &
\multicolumn{1}{c}{--} & \multicolumn{1}{c}{--} & 0.007 & 0.000 &
$-$0.002 & 1.03 & 100 & f & 1/20\\ 4 & 60.0 & 40.1 & 61.0 & 346.4 &
\multicolumn{1}{c}{--} & \multicolumn{1}{c}{--} &
\multicolumn{1}{c}{--} & 0.002 & $-$0.002 & 0.002 & 0.94 & 20 & f &
6/20\\

5 & 63.0 & 40.0 & 57.4 & 340.7 & $-$0.1 & 85.7 & 95.7 &
\multicolumn{1}{c}{--} & \multicolumn{1}{c}{--} &
\multicolumn{1}{c}{--} & 0.99 & 100 & fp & 7/20\\ 6 & 70.6 & 40.2 &
49.9 & 345.2 & $-$1.3 & 5.2 & 241.5 & \multicolumn{1}{c}{--} &
\multicolumn{1}{c}{--} & \multicolumn{1}{c}{--} & 1.01 & 20 & fp &
13/20\\

7 & 62.3 & 39.9 & 57.1 & 339.7 & 0.0 & 23.2 & 265.3 &
\multicolumn{1}{c}{--} & \multicolumn{1}{c}{--} &
\multicolumn{1}{c}{--} & 1.02 & 100 & f & 2/6\\ 8 & 45.5 & 40.0 & 66.0
& 358.0 & 0.5 & 23.5 & 228.6 & \multicolumn{1}{c}{--} &
\multicolumn{1}{c}{--} & \multicolumn{1}{c}{--} & 1.00 & 20 & f &
2/6\\

9 & 56.2 & 39.8 & 63.9 & 339.6 & $-$2.6 & 31.6 & 249.0 & $-$0.013 &
$-$0.007 & $-$0.004 & 1.02 & 100 & fp & 1/20\\
10 & 45.7 & 39.7 & 56.2 & 353.4 & $-$20.9 & 38.8 & 179.8 & 0.075 &
0.001 & $-$0.085 & 1.02 & 20 & f & 3/20\\[0.8ex] 
 

\multicolumn{15}{l}{\itshape (A2), centred dipole viewed at
$i=30\degr$:}\\ & \textbf{30.0} & \textbf{40.0} & \textbf{60.0} &
\textbf{340.0} & \textbf{0.0} & \textbf{0.0} & \textbf{0.0} &
\textbf{0.0} & \textbf{0.0} & \textbf{0.0} & \multicolumn{1}{c}{--} &
\multicolumn{1}{c}{--} & \multicolumn{1}{c}{--} &
\multicolumn{1}{c}{--}\\

1 & 25.6 & 41.5 & 58.1 & 338.6 & \multicolumn{1}{c}{--} &
\multicolumn{1}{c}{--} & \multicolumn{1}{c}{--} & 0.018 & 0.003 &
$-$0.014 & 1.03 & 100 & fp & 8/20\\ 2 & 29.0 & 39.9 & 56.8 & 345.9 &
\multicolumn{1}{c}{--} & \multicolumn{1}{c}{--} &
\multicolumn{1}{c}{--} & 0.000 & $-$0.006 & $-$0.001 & 1.00 & 20 & fp
& 9/20\\

3 & 29.5 & 39.5 & 59.0 & 339.6 & $-$7.4 & 22.0 & 300.6 & $-$0.068 &
$-$0.024 & 0.020 & 1.02 & 100 & fp & 1/20\\[0.8ex]


\multicolumn{15}{l}{\itshape (B), centred quadrupole viewed at
$i=60\degr$:}\\ & \textbf{60.0} & \textbf{0.0} & \textbf{60.0} &
\textbf{340.0} & \textbf{40.0} & \textbf{60.0} & \textbf{340.0} &
\textbf{0.0} & \textbf{0.0} & \textbf{0.0} & \multicolumn{1}{c}{--} &
\multicolumn{1}{c}{--} & \multicolumn{1}{c}{--} & \multicolumn{1}{c}{-
-}\\

1 & 56.7 & 0.1 & 63.0 & 339.9 & 40.0 & 63.0 & 339.9 &
\multicolumn{1}{c}{--} & \multicolumn{1}{c}{--} &
\multicolumn{1}{c}{--} & 0.97 & 100 & fp & 5/10\\ 2 & 58.6 & $-$0.8 &
60.8 & 331.2 & 40.0 & 60.8 & 331.2 & \multicolumn{1}{c}{--} &
\multicolumn{1}{c}{--} & \multicolumn{1}{c}{--} & 1.02 & 20 & fp &
5/10\\

3 & 63.7 & 0.0 & 56.6 & 339.3 & 40.0 & 56.6 & 339.3 &
\multicolumn{1}{c}{--} & \multicolumn{1}{c}{--} &
\multicolumn{1}{c}{--} & 1.00 & 100 & f & 2/20\\ 4 & 49.1 & $-$0.4 &
61.7 & 346.2 & 40.0 & 61.7 & 346.2 & \multicolumn{1}{c}{--} &
\multicolumn{1}{c}{--} & \multicolumn{1}{c}{--} & 1.00 & 20 & f &
5/20\\ [0.8ex]


\multicolumn{15}{l}{\itshape (C), aligned centred dipole and
quadrupole viewed at $i=60\degr$:}\\ & \textbf{60.0} & \textbf{40.0} &
\textbf{60.0} & \textbf{340.0} & \textbf{20.0} & \textbf{60.0} &
\textbf{340.0} & \textbf{0.0} & \textbf{0.0} & \textbf{0.0} &
\multicolumn{1}{c}{--} & \multicolumn{1}{c}{--} &
\multicolumn{1}{c}{--} & \multicolumn{1}{c} {--}\\

1 & 58.8 & 40.0 & 61.3 & 340.1 & 20.1 & 61.3 & 340.1 &
\multicolumn{1}{c}{--} & \multicolumn{1}{c}{--} &
\multicolumn{1}{c}{--} & 0.93 & 100 & fp & 5/10\\ 2 & 55.2 & 40.1 &
64.4 & 334.7 & 19.8 & 64.4 & 334.7 & \multicolumn{1}{c}{--} &
\multicolumn{1}{c}{--} & \multicolumn{1}{c}{--} & 1.00 & 20 & fp &
5/10\\

3 & 58.5 & 40.0 & 64.8 & 340.6 & 20.1 & 64.8 & 340.6 &
\multicolumn{1}{c}{--} & \multicolumn{1}{c}{--} &
\multicolumn{1}{c}{--} & 1.05 & 100 & f & 3/20\\ 4 & 51.1 & 40.0 &
65.4 & 338.3 & 20.0 & 65.4 & 338.3 & \multicolumn{1}{c}{--} &
\multicolumn{1}{c}{--} & \multicolumn{1}{c}{--} & 0.99 & 20 & f &
9/20\\ [0.8ex] 


\multicolumn{15}{l}{\itshape (D), shifted high-field dipole viewed at
$i=60\degr$:}\\ & \textbf{60.0} & \textbf{110.0} & \textbf{0.0} &
\textbf{0.0} & \textbf{0.0} & \textbf{0.0} & \textbf{0.0} &
\textbf{0.05} & \textbf{$-$0.10} & \textbf{0.15} &
\multicolumn{1}{c}{--} & \multicolumn{1}{c}{--} &
\multicolumn{1}{c}{--} & \multicolumn{1}{c} {--}\\

1 & 60.1 & 110.1 & 0.3 & 150.2 & \multicolumn{1}{c}{--} &
\multicolumn{1}{c}{--} & \multicolumn{1}{c}{--} & 0.051 & $-$0.101 &
0.149 & 1.01 & 100 & fp & 1/10\\ 2 & 61.1 & 109.7 & 0.3 & 223.6 &
\multicolumn{1}{c}{--} & \multicolumn{1}{c}{--} &
\multicolumn{1}{c}{--} & 0.056 & $-$0.101 & 0.147 & 1.00 & 20 & fp &
3/10\\

3 & 59.8 & 109.9 & 0.1 & 218.1 & \multicolumn{1}{c}{--} &
\multicolumn{1}{c}{--} & \multicolumn{1}{c}{--} & 0.050 & $-$0.099 &
0.150 & 0.97 & 100 & f & 6/20\\ 4 & 59.0 & 110.4 & 2.4 & 160.1 &
\multicolumn{1}{c}{--} & \multicolumn{1}{c}{--} &
\multicolumn{1}{c}{--} & 0.045 & $-$0.105 & 0.154 & 1.05 & 20 & f &
1/20\\

5 & 59.8 & 109.8 & 0.1 & 90.1 & $-$0.7 & 81.1 & 266.9 & 0.050 &
$-$0.100 & 0.150 & 0.99 & 100 & fp & 2/20\\ 

6 & 58.8 & 110.0 & 1.0 & 258.8 & 6.0 & 67.4 & 165.5 & 0.053 & $-$0.105
& 0.162 & 0.93 & 20 & f & 18/20\\ [0.8ex]

 
\multicolumn{15}{l}{\itshape (E), shifted dipole viewed at
$i=60\degr$:}\\ & \textbf{60.0} & \textbf{58.0} & \textbf{60.0} &
\textbf{340.0} & \textbf{0.0} & \textbf{0.0} & \textbf{0.0} &
\textbf{0.15} & \textbf{$-$0.10} & \textbf{0.30} &
\multicolumn{1}{c}{--} & \multicolumn{1}{c}{--} &
\multicolumn{1}{c}{--} & \multicolumn{1}{ c}{--}\\

1 & 63.9 & 57.4 & 59.1 & 338.3 & \multicolumn{1}{c}{--} &
\multicolumn{1}{c}{--} & \multicolumn{1}{c}{--} & 0.15 & $-$0.09 &
0.30 & 1.02 & 100 & fp & 4/10\\ 2 & 60.4 & 57.9 & 60.9 & 337.0 &
\multicolumn{1}{c}{--} & \multicolumn{1}{c}{--} &
\multicolumn{1}{c}{--} & 0.16 & $-$0.08 & 0.30 & 1.05 & 20 & fp &
3/10\\

3 & 60.2 & 57.6 & 60.3 & 340.3 & \multicolumn{1}{c}{--} &
\multicolumn{1}{c}{--} & \multicolumn{1}{c}{--} & 0.15 & $-$0.10 &
0.30 & 1.04 & 100 & f & 10/20\\ 4 & 56.4 & 59.1 & 61.3 & 341.0 &
\multicolumn{1}{c}{--} & \multicolumn{1}{c}{--} &
\multicolumn{1}{c}{--} & 0.16 & $-$0.10 & 0.30 & 0.97 & 20 & f &
15/20\\

5 & 60.1 & 59.5 & 63.9 & 340.5 & 15.2 & 73.1 & 334.1 & 0.13 & $-$0.09
& 0.23 & 1.06 & 100 & fp & 5/20\\

6 & 54.4 & 62.9 & 61.3 & 355.2 & 26.7 & 51.7 & 256.3 & 0.22 & $-$0.16
& 0.29 & 1.00 & 20 & f & 11/20\\ [0.8ex]


\multicolumn{15}{l}{\itshape (F), non-aligned dipole-quadrupole
combination viewed at $i=60\degr$:}\\ 

& \textbf{60.0} & \textbf{40.0} & \textbf{60.0} & \textbf{340.0} &
\textbf{40.0} & \textbf{30.0} & \textbf{250.0} & \textbf{0.0} &
\textbf{0.0} & \textbf{0.0} & \multicolumn{1}{c}{--} &
\multicolumn{1}{c}{--} & \multicolumn{1}{c}{--} & \multicolumn{1}{c}
{--}\\

1 & 58.8 & 40.3 & 58.5 & 337.3 & 39.8 & 31.9 & 247.3 &
\multicolumn{1}{c}{--} & \multicolumn{1}{c}{--} &
\multicolumn{1}{c}{--} & 1.01 & 100 & fp & 4/20\\

2 & 58.6 & 39.6 & 51.1 & 350.1 & 41.4 & 26.6 & 233.2 &
\multicolumn{1}{c}{--} & \multicolumn{1}{c}{--} &
\multicolumn{1}{c}{--} & 1.05 & 20 & fp & 10/20\\ 

3 & 66.2 & 39.9 & 68.0 & 347.4 & 40.4 & 26.7 & 270.0 &
\multicolumn{1}{c}{--} & \multicolumn{1}{c}{--} &
\multicolumn{1}{c}{--} & 1.13 & 100 & f & 5/20\\ 

4 & 43.3 & 49.1 & 51.2 & 273.6 & $-$32.2 & 66.1 & 329.8 &
\multicolumn{1}{c}{--} & \multicolumn{1}{c}{--} &
\multicolumn{1}{c}{--} & 0.94 & 20 & f & 10/20\\

5 & 55.4 & 38.6 & 61.5 & 340.6 & 39.2 & 36.4 & 251.6 & $-$0.022 &
0.009 & $-$0.002 & 1.05 & 100 & fp & 1/20\\

6 & 46.7 & 39.4 & 36.7 & 269.9 & 30.4 & 39.0 & 248.8 & 0.051 & 0.078 &
$-$0.068 & 1.05 & 20 & f & 1/20\\ 

\noalign{\smallskip} \hline

\end{tabular}
\end{table*}

\begin{figure*}[t]
\includegraphics[width=18.0cm]{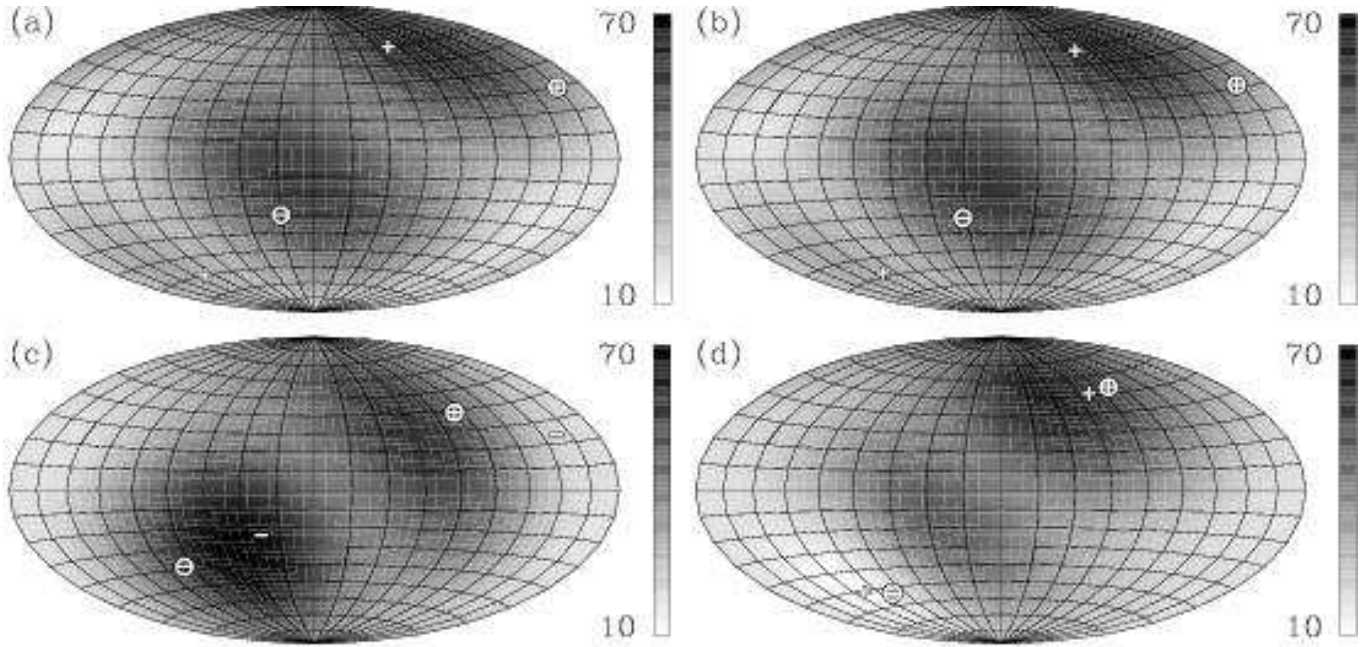}
\caption{Aitoff maps of the field strength $B$ (in MG) showing the
results of reconstruction fits for the Model (F) field distribution
(non-aligned dipole-quadrupole combination). The centre of the maps
marks the $-x_0$ direction. The axis of the quadrupole and the field
direction at its poles is indicated by the $+$ and $-$ symbols, the
axis of the dipole correspondingly by the circled $+$ and $-$ symbols
(see text). \textbf{a)} Input field, \textbf{b)} low-noise
flux-and-polarization reconstruction of Table~\ref{tab:input}, Model
(F), line~1, \textbf{c)} high-noise flux-only reconstruction of
line~4, and \textbf{d)} high-noise flux-only reconstruction of line~6.}
\label{fig:reconstruction_F}
\end{figure*}

{\itshape Model (F), non-aligned dipole-quadrupole combination viewed
at $i=60\degr$:} All fits reproduce the general structure of the field
with its two spots, separated by $\sim$90\degr, but only the fit using
flux and polarization spectra at \mbox{$S/N=100$} (line~1) correctly
finds the axes of both, the dipole and the quadrupole components.
Convergence problems, which arise when the polarization is disregarded
(line~3), may be due to a more corrugated \chisq-landscape compared
with the fits including the polarization.
Figure~\ref{fig:reconstruction_F} provides an overview of the results
for model (F). The input is depicted in
Fig.~\ref{fig:reconstruction_F}a and the line-1 reconstruction in
Fig.~\ref{fig:reconstruction_F}b. The high-noise flux-only fits of
lines~4 and 6 (Figs.~\ref{fig:reconstruction_F}c and
\ref{fig:reconstruction_F}d) deviate in the strengths and orientations
of the dipoles and quadrupoles. In Figs.~\ref{fig:reconstruction_F}a
and \ref{fig:reconstruction_F}b, the high-field spot on the northern
(southern) hemisphere is dominated by the quadrupolar (dipolar)
contribution.  In Fig.~\ref{fig:reconstruction_F}c, the polarity of
the quadrupole is reversed. Finally, in
Fig.~\ref{fig:reconstruction_F}d, dipole and quadrupole are nearly
aligned, but the spots are shifted due to the finite values of \xoff\
and \yoff. Nevertheless, these fits are not altogether wrong if only
the two-spot structure is considered.  They suggest that a low
\chisqred\ does, in fact, indicate a representation which bears some
similarity to the input field structure, even if the choice of
multipoles differs from that of the input. Seemingly, better results
cannot be expected given the high noise of these two fits.

In summary, the code is able to reconstruct the magnetic field
geometries of the type discussed here from phase-resolved flux and
polarization spectra of high $S/N$. Our experience is that the
inclusion of more than four phases does not improve the fits
substantially, which is understandable given the overlap in surface
coverage. Naturally, the reconstruction becomes less perfect when the
polarization information is excluded and the noise is
increased. Leaving off the polarization also seems to create
convergence problems. An important aspect is that phase-dependent
spectra allow the inclination of the rotation axis to be determined
along with the field geometry. The accuracy achieved ranges from a few
degrees to about 20\degr\ depending on the quality of the spectra.

\subsubsection{Fits to a single spectrum}
If only a single set of flux and polarization spectra or a single flux
spectrum is available, information on the field distribution is
reduced to the visible hemisphere. The location of the rotation axis
remains unknown, and only the angle between the magnetic axis and the
line of sight is constrained by the fit. We have performed similar
tests as above to single sets of spectra and find that simple field
geometries can still be recovered.

\section{Discussion}
\label{sec:discussion}

We have presented a formalized approach to the interpretation of
phase-resolved flux and circular polarization spectra of rotating
magnetic white dwarfs (MWDs).  Tomographically locating positions with
a certain field strength $\vec{B}$ on the surface of the star is
hampered by the fact that only the self-eclipse of such a region
manifests itself in flux spectra, while the positional information
contributed by the rotation is obliterated by the Stark broadening.

\subsection{Present approach}

In our approach, we determine the parameters of a global field model
directly by a least-squares fit to the spectral data. We caution that
it is not \emph{a priori} clear to what extent the global field can be
constrained by such an approach, because the spectral information
represents an average over the visible hemisphere at each phase.  Our
results demonstrate, however, that the phase-resolved Zeeman spectra
contain enough information to allow the reconstruction of the field
geometries considered by us. These involve combinations of dipoles and
quadrupoles which are allowed to have different axes and to be shifted
off-centre. The model contains up to ten free parameters and is
sufficiently general to allow for rather complex surface field
geometries featuring, e.g., a dominant single high-field spot, two
spots separated by much less than 180\degr, or even a bipolar spot on
an otherwise low-field star (not included in the models presented
here). An advantage of our approach is that these fields automatically
fulfil the requirement of being produced by sources inside the star. A
disadvantage is the limitation in the number of free parameters.

In addition to the cases presented here, we have also attempted to
reconstruct octupolar fields and were successful for aligned
dipole-quadrupole-octupole combinations. However, if all multipole
components with \mbox{$l=3$} and \mbox{$m=0,\dots,3$} are included (15
parameters for the expansion, two angles describing the direction of
the reference axis, and the inclination), the \texttt{evoC}
minimization algorithm encounters convergence problems, caused by too
large a number of free parameters.

Another important aspect is the level of the signal-to-noise ratio
$S/N$ required for a successful reconstruction of the field. The model
atmospheres of hydrogen-rich MWDs are characterized by the rather
broad and strong Zeeman-shifted Balmer lines which allow a field
reconstruction already for \mbox{$S/N=20$--100}: \mbox{$S/N=20$} is
the lower limit, while there is little improvement for
\mbox{$S/N>100$}. For comparison, Zeeman-Doppler imaging of main
sequence stars operates on much fainter metal lines and needs a much
higher $S/N$ \citep{brownetal91-1}. However, while Zeeman-Doppler
imaging is performed over individual lines, the high field strengths
of the MWDs require a fit over the whole visible wavelength range.

In the analysis of observed Zeeman spectra, one may encounter some
problems which are absent in the present reconstruction of synthetic
spectra.  While the variation of the statistical noise amplitude with
wavelength can be accounted for in the \chisq-statistic
(Eq.~\ref{eq:chisq}), systematic uncertainties between the observed
and calculated spectra cannot: (i) errors in the theoretical database
spectra; and (ii) errors in the flux calibration of the observed
spectra.  Errors of type (i) may prevent a satisfactory convergence of
the fits and/or lead to incorrect values of the parameters describing
the field. Our experience is that such errors are of minor importance,
given the present state of the theory of radiative transfer in
magnetic stellar atmospheres.  Errors of type (ii) may affect the
ability to recognize high field spots on stars with a predominantly
moderate field. For example, in model (F) at phase \mbox{$\phi=0.25$},
and similarly in other models, the H$\alpha$~$\sigma^-$ component
consists of a shallow depression extending from 5000 to 6000\,\AA. An
error in the flux calibration which happens to weaken or strengthen
such a depression can lead to serious errors in the derived field
distribution. A careful flux calibration is, therefore, of utmost
importance.

\subsection{Different optimization strategies}

Should MWDs turn out to have field geometries which are more complex
and require more free parameters than adopted by us, we may have to
consider alternative optimization techniques. E.g., a genetic
algorithm may be more robust than the \texttt{evoC} code and allow for
a somewhat larger number of parameters. A full harmonic expansion with
\mbox{$l\gg 2$} may become tractable if a regularization operator like
MEM drives the solution towards low-order fields while permitting
higher-order components to be used as necessary to fit the data.

\subsection{The ZEBRA approach}

The problem of retrieving the field structure of rotating MWDs has
previously been studied by \citet{donatietal94-1} in what they called
the ZEBRA approach. They used a maximum-entropy method to deduce the
most likely two-dimensional frequency distribution
$f(B_\mathrm{t},B_\mathrm{l})$ of the transverse and longitudinal
field components with respect to the line of sight separately for each
rotational phase. The method has the obvious advantage that no \emph{a
priori} assumption is made about the global field structure. On the
other hand, the interrelation between the overlapping field
distributions at different rotational phases is not utilized and there
is no prescription for the interpretation of such an interrelation in
terms of a global field. Indeed, there is no guarantee of a physically
meaningful reconstruction (e.g. sources only within the star leading
to a curl-free field outside the star). Thus, the detailed structure
and the physical characteristics of the global field remain undefined
in the ZEBRA method in the present form \citep{donatietal94-1}.

If the underlying global field structure is sufficiently simple, it
may be derived in a second step added to the ZEBRA method. In a first
step, the best-fit ZEBRA diagrams (or {\bpd}s similar to ours) are
determined using a MEM-type regularization scheme as suggested by
\citet{donatietal94-1}. In a second step, a parametrized global
magnetic field model is then fitted to the phase-resolved ZEBRA
diagrams. Since the second step would not involve the computation of
spectra from the database, which is by far the most time-critical
process in the present method, this two-step approach is probably
advantageous with respect to CPU time. Without detailed tests,
however, it is not clear whether this approach would be superior to
directly fitting the Zeeman spectra.

\subsection{A future approach}

One may endeavour to relax the restrictions on the global field
structure by parametrizing the surface field as
\mbox{$(\vec{B}_1,\dots,\vec{B}_N)$} for a star with $N$ surface
elements and to impose a regularization scheme, e.g. MEM, to ensure
the smoothness of the solution.  The feasibility of such an approach,
its convergence properties, and the interpretation of the derived
field model would still have to be studied, however, as well as the
demands on computation time given the formidable number of parameters.

\section{Conclusion}

We have described a method to reconstruct the field structure of
magnetic white dwarfs which provides an internally consistent fit to
spectropolarimetric data taken at different rotational phases in terms
of a parametrized field model.  We presently use dipoles and
quadrupoles which are allowed to be shifted off-centre to increase the
versatility of the model.  An application to real data will be
described elsewhere.

We do not know whether MWDs have the regular fields adopted here or
possibly field structures as complex as spotted main sequence stars.
Fortunately, there are several single white dwarfs with known
rotational periods, and about 60 rotating MWDs in cataclysmic
binaries, some of which are known to have fields which deviate from
simple centred dipoles. The study of such systems using the present
and similar techniques promises to increase our knowledge of the
end-product of magnetic stellar evolution.

\begin{acknowledgements}
We thank Torsten Rahn for his contributions to the Kiel fraction of
this program and Klaus Reinsch for fruitful discussions. FE acknowledges
a grant from Graduiertenf\"orderung des Landes Niedersachsen. This
work was supported in part by BMBF/DLR grant 50\,OR\,9903\,6.
\end{acknowledgements}
\bibliographystyle{aa}
\bibliography{aamnem99,mylit}

\begin{thebibliography}{36}
\expandafter\ifx\csname natexlab\endcsname\relax\def\natexlab#1{#1}\fi

\bibitem[{{Beckers}(1969)}]{beckers69-1}
{Beckers}, J.~M. 1969, Sol. Phys., 9, 372

\bibitem[{{Brown} {et~al.}(1991){Brown}, {Donati}, {Rees}, \&
  {Semel}}]{brownetal91-1}
{Brown}, S.~F., {Donati}, J.-F., {Rees}, D.~E., \& {Semel}, M. 1991, A\&A, 250,
  463

\bibitem[{{Burleigh} {et~al.}(1999){Burleigh}, {Jordan}, \&
  {Schweizer}}]{burleighetal99-1}
{Burleigh}, M.~R., {Jordan}, S., \& {Schweizer}, W. 1999, ApJ Lett., 510, L37

\bibitem[{{Dittmann}(1995)}]{dittmann95-1}
{Dittmann}, O. 1995, Dissertation, Heidelberg

\bibitem[{{Donati} {et~al.}(1994){Donati}, {Achilleos}, {Matthews}, \&
  {Wesemael}}]{donatietal94-1}
{Donati}, J.-F., {Achilleos}, N., {Matthews}, J.~M., \& {Wesemael}, F. 1994,
  A\&A, 285, 285

\bibitem[{{Donati} {et~al.}(1989){Donati}, {Semel}, \&
  {Praderie}}]{donatietal89-1}
{Donati}, J.-F., {Semel}, M., \& {Praderie}, F. 1989, A\&A, 225, 467

\bibitem[{{Euchner}(1998)}]{euchner98-1}
{Euchner}, F. 1998, Diplomarbeit, Georg-August-Universit{\"a}t G{\"o}ttingen

\bibitem[{{Forster} {et~al.}(1984){Forster}, {Strupat}, {R{\"o}sner}, {Wunner},
  {Ruder}, \& {Herold}}]{forsteretal84-1}
{Forster}, H., {Strupat}, W., {R{\"o}sner}, W., {et~al.} 1984, J. Phys. V, 17,
  1301

\bibitem[{{G{\"a}nsicke} \& {Beuermann}(1996)}]{gaensicke+beuermann96-1}
{G{\"a}nsicke}, B.~T. \& {Beuermann}, K. 1996, A\&A, 309, L47

\bibitem[{{G{\"a}nsicke} {et~al.}(1998){G{\"a}nsicke}, {Hoard}, {Beuermann},
  {Sion}, \& {Szkody}}]{gaensickeetal98-1}
{G{\"a}nsicke}, B.~T., {Hoard}, D.~W., {Beuermann}, K., {Sion}, E.~M., \&
  {Szkody}, P. 1998, A\&A, 338, 933

\bibitem[{{Hardorp} {et~al.}(1976){Hardorp}, {Shore}, \&
  {Wittmann}}]{hardorpetal76-1}
{Hardorp}, J., {Shore}, S.~N., \& {Wittmann}, A. 1976, in Physics of Ap stars,
  ed. W.~W. {Weiss}, H.~{Jenkner}, \& H.~J. {Wood}, IAU Coll. No.~32, 419

\bibitem[{{Jordan}(1992)}]{jordan92-1}
{Jordan}, S. 1992, A\&A, 265, 570

\bibitem[{{Jordan}(2001)}]{jordan01-1}
{Jordan}, S. 2001, in 12th European Workshop on White Dwarfs, ed. J.~L.
  {Provencal}, H.~L. {Shipman}, J.~{MacDonald}, \& S.~{Goodchild}, ASP Conf.\
  Ser. No. 226 (San Francisco: Astronomical Society of the Pacific), 269

\bibitem[{{Jordan} \& {Merani}(1995)}]{jordan+merani95-1}
{Jordan}, S. \& {Merani}, N. 1995, in 9th European Workshop on White Dwarfs,
  ed. D.~{Koester} \& K.~{Werner}, Lecture Notes in Physics No. 443 (Berlin:
  Springer Verlag), 134

\bibitem[{{Koester} {et~al.}(1979){Koester}, {Schulz}, \&
  {Weidemann}}]{koesteretal79-1}
{Koester}, D., {Schulz}, H., \& {Weidemann}, V. 1979, A\&A, 76, 262

\bibitem[{{Kube} {et~al.}(2000){Kube}, {G{\" a}nsicke}, \&
  {Beuermann}}]{kubeetal00-1}
{Kube}, J., {G{\" a}nsicke}, B.~T., \& {Beuermann}, K. 2000, A\&A, 356, 490

\bibitem[{{Lamb} \& {Sutherland}(1974)}]{lamb+sutherland74-1}
{Lamb}, F.~K. \& {Sutherland}, P.~G. 1974, in Physics of Dense Matter, ed.
  C.~J. {Hansen}, IAU Symp. No.~53 (Dordrecht: Reidel), 265

\bibitem[{{Langel}(1987)}]{langel87-1}
{Langel}, R.~A. 1987, in Geomagnetism, ed. J.~A. {Jacobs}, Vol.~1 (London:
  Academic Press), 249

\bibitem[{{Merani} {et~al.}(1995){Merani}, {Main}, \&
  {Wunner}}]{meranietal95-1}
{Merani}, N., {Main}, J., \& {Wunner}, G. 1995, in Astrophysical Applications
  of Powerful New Databases, ed. S.~J. {Adelman} \& W.~L. {Wiese}, ASP Conf.\
  Ser. No.~78 (San Francisco: Astronomical Society of the Pacific), 81

\bibitem[{{Muslimov} {et~al.}(1995){Muslimov}, {Van Horn}, \&
  {Wood}}]{muslimovetal95-1}
{Muslimov}, A.~G., {Van Horn}, H.~M., \& {Wood}, M.~A. 1995, ApJ, 442, 758

\bibitem[{{Nelder} \& {Mead}(1965)}]{nelder+mead65-1}
{Nelder}, J.~A. \& {Mead}, R. 1965, Computer Journal, 7, 308

\bibitem[{{Press} {et~al.}(1992){Press}, {Teukolsky}, {Vetterling}, \&
  {Flannery}}]{pressetal92-1}
{Press}, W.~H., {Teukolsky}, S.~A., {Vetterling}, W.~T., \& {Flannery}, B.~P.
  1992, Numerical Recipes in C, 2nd edn. (Cambridge University Press)

\bibitem[{{Putney} \& {Jordan}(1995)}]{putney+jordan95-1}
{Putney}, A. \& {Jordan}, S. 1995, ApJ, 449, 863

\bibitem[{{Rahn}(1999)}]{rahn99-1}
{Rahn}, T. 1999, Diplomarbeit, Christian-Albrechts-Universit{\"a}t Kiel

\bibitem[{{Ramaty}(1969)}]{ramaty69-1}
{Ramaty}, R. 1969, ApJ, 158, 753

\bibitem[{{Rechenberg}(1994)}]{rechenberg94-1}
{Rechenberg}, I. 1994, Evolutionsstrategie '94, Werkstatt Bionik und
  Evolutionstechnik No.~1 (Stuttgart: frommann-holzboog)

\bibitem[{{R{\"o}sner} {et~al.}(1984){R{\"o}sner}, {Wunner}, {Herold}, \&
  {Ruder}}]{roesneretal84-1}
{R{\"o}sner}, W., {Wunner}, G., {Herold}, H., \& {Ruder}, H. 1984, J. Phys. V,
  17, 29

\bibitem[{{Schwope}(1995)}]{schwope95-1}
{Schwope}, A.~D. 1995, Rev. Mod. Astron., 8, 125

\bibitem[{{Semel}(1989)}]{semel89-1}
{Semel}, M. 1989, A\&A, 225, 456

\bibitem[{{Takeda}(1991)}]{takeda91-1}
{Takeda}, Y. 1991, PASJ, 43, 719

\bibitem[{{Trint} \& {Utecht}(1994)}]{trint+utecht94-1}
{Trint}, K. \& {Utecht}, U. 1994,
  ftp://ftp-bionik.fb10.tu-berlin.de/pub/software/evoC/

\bibitem[{{Unno}(1956)}]{unno56-1}
{Unno}, W. 1956, PASJ, 8, 108

\bibitem[{{Wickramasinghe} \& {Cropper}(1988)}]{wickramasinghe+cropper88-1}
{Wickramasinghe}, D.~T. \& {Cropper}, M. 1988, MNRAS, 235, 1451

\bibitem[{{Wickramasinghe} \& {Ferrario}(2000)}]{wickramasinghe+ferrario00-1}
{Wickramasinghe}, D.~T. \& {Ferrario}, L. 2000, PASP, 112, 873

\bibitem[{{Wickramasinghe} \& {Martin}(1979)}]{wickramasinghe+martin79-1}
{Wickramasinghe}, D.~T. \& {Martin}, B. 1979, MNRAS, 189, 883

\bibitem[{{Wunner} {et~al.}(1985){Wunner}, {R{\"o}sner}, {Herold}, \&
  {Ruder}}]{wunneretal85-1}
{Wunner}, G., {R{\"o}sner}, W., {Herold}, H., \& {Ruder}, H. 1985, A\&A, 149,
  102

\end{thebibliography}

\end{document}